\documentclass[10pt,3p]{elsarticle}

\usepackage{stix2}
\usepackage{amsmath}
\usepackage{amsthm}
\usepackage{graphicx}
\usepackage{epstopdf} 
\usepackage{bm}
\usepackage{todonotes}
\usepackage{booktabs}
\usepackage{caption}
\usepackage{subcaption}
\usepackage{hyperref}
\usepackage{placeins}
\usepackage[mathlines]{lineno}
\usepackage{comment}
\usepackage{cancel}

\theoremstyle{plain}

\theoremstyle{plain}
\newtheorem{remark}{Remark}

\theoremstyle{definition}

\newcommand{\rd}{\mathrm{d}}
\newcommand{\vt}[1]{\bm{#1}}

\newcommand{\dd}[2]{\frac{\partial #1}{\partial #2}}
\newcommand{\Pra}{\mathrm{Pr}}
\newcommand{\Ra}{\mathrm{Ra}}
\newcommand{\Ge}{\mathrm{Ge}}
\newcommand{\Nu}{\mathrm{Nu}}
\newcommand{\uref}{u_{\mathrm{ref}}}
\newcommand{\ek}{e_k}
\newcommand{\eint}{e_i}

\makeatletter
\def\ps@pprintTitle{%
 \let\@oddhead\@empty
 \let\@evenhead\@empty
 \def\@oddfoot{}%
 \let\@evenfoot\@oddfoot}
\makeatother

\newcommand*\linenomathpatchAMS[1]{%
  \expandafter\pretocmd\csname #1\endcsname {\linenomathAMS}{}{}%
  \expandafter\pretocmd\csname #1*\endcsname{\linenomathAMS}{}{}%
  \expandafter\apptocmd\csname end#1\endcsname {\endlinenomath}{}{}%
  \expandafter\apptocmd\csname end#1*\endcsname{\endlinenomath}{}{}%
}

\expandafter\ifx\linenomath\linenomathWithnumbers
  \let\linenomathAMS\linenomathWithnumbers
  \patchcmd\linenomathAMS{\advance\postdisplaypenalty\linenopenalty}{}{}{}
\else
  \let\linenomathAMS\linenomathNonumbers
\fi

\linenomathpatchAMS{gather}
\linenomathpatchAMS{multline}
\linenomathpatchAMS{align}
\linenomathpatchAMS{alignat}
\linenomathpatchAMS{flalign}

\begin{document}

\begin{frontmatter}

\title{Energy-consistent discretization of viscous dissipation with application to natural convection flow}

\author[1]{B. Sanderse}
\author[2]{F.X. Trias}

\affiliation[1]{organization={Centrum Wiskunde \& Informatica},%Department and Organization
            addressline={Science Park 123}, 
            city={Amsterdam},
            country={The Netherlands}}
\affiliation[2]{organization={Heat and Mass Transfer Technological Center, Technical University of Catalonia, ESEIAAT}, 
addressline={c/ Colom 11, 08222 Terrassa (Barcelona)},
country={Spain}}

\begin{abstract}
A new energy-consistent discretization of the viscous dissipation function in incompressible flows is proposed. It is \textit{implied} by choosing a discretization of the diffusive terms and a discretization of the local kinetic energy equation and by requiring that continuous identities like the product rule are mimicked discretely. The proposed viscous dissipation function has a quadratic, strictly dissipative form, for both simplified (constant viscosity) stress tensors and general stress tensors. The proposed expression is not only useful in evaluating energy budgets in turbulent flows, but also in natural convection flows, where it appears in the internal energy equation and is responsible for viscous heating. The viscous dissipation function is such that a \textit{consistent total energy balance} is obtained: the `implied' presence as sink in the kinetic energy equation is exactly balanced by explicitly adding it as source term in the internal energy equation.

Numerical experiments of Rayleigh-Bénard convection (RBC) and Rayleigh-Taylor instabilities confirm that with the proposed dissipation function, the energy exchange between kinetic and internal energy is exactly preserved. The experiments show furthermore that viscous dissipation does not affect the critical Rayleigh number at which instabilities form, but it does significantly impact the development of instabilities once they occur. Consequently, the value of the Nusselt number on the cold plate becomes larger than on the hot plate, with the difference increasing with increasing Gebhart number. Finally, 3D simulations of turbulent RBC show that energy balances are exactly satisfied even for very coarse grids; therefore, we consider that the proposed discretization forms an excellent starting point for testing sub-grid scale models.

\end{abstract}


\end{frontmatter}
    
\section{Introduction and problem description}
In this article we study the viscous dissipation function and its role in natural convection flows described by the incompressible Navier-Stokes equations, with buoyancy effects modelled by the Boussinesq approximation \cite{siggia1994}. These `Boussinesq` or `Oberbeck-Boussinesq' equations have attracted much scientific interest over several decades \cite{batchelor1954}, not only because of their physical relevance, but also of their intriguing mathematical properties. An important test case studied with the Boussinesq system is that of Rayleigh-Bénard convection \cite{grossmann2000}, in which a box of fluid is heated from the bottom and cooled from the top, giving rise to convection cells. The Boussinesq equations also describe a (miscible) form of Rayleigh-Taylor instability, which occurs when a heavy (cold) fluid is positioned above a light (warm) fluid.

A common assumption in many incompressible natural convection studies is that the effect of viscous dissipation on the internal energy (effectively on the temperature) is neglected. This assumption is not always valid, for example when considering natural convection in the Earth mantle, when considering highly viscous liquids, when large length scales are involved, or in devices operating at high rotational speed \cite{barenblatt2003,mckenzie1974,schubert2001,kee2003,hewitt1975,turcotte1974,gebhart1962,ostrach1952}. Of course, when considering compressible flows, e.g.\ high-speed flows, including heating by viscous dissipation is known to be important, and several benchmarking studies have been performed related to modelling natural convection in the Earth mantle \cite{blankenbach1989,king2010}. These studies typically assume infinite Prandtl numbers, and ignore the unsteady and convective terms in the momentum equations. In this paper we will restrict ourselves to the incompressible situation, for which this effect is less studied. In this incompressible case, Ostrach \cite{ostrach1952}, Gebhart \cite{gebhart1962} and Turcotte et al.\ \cite{turcotte1974} should be explicitly mentioned, being among the first to address the role of viscous dissipation and to introduce next to the well-known Rayleigh and Prandtl numbers another dimensionless quantity, which is known as the dissipation number or the Gebhart number. In addition, we mention the work of Barletta and co-authors \cite{barletta2008,barletta2009a,barletta2011,barletta2022a}, who considered the role of viscous dissipation in natural convection in several papers, studying the correct mathematical formulation of the problem and linear stability analysis for different geometries.  Turcotte et al.\ \cite{turcotte1974} were probably one of the first to perform numerical experiments of incompressible natural convection flows that include viscous dissipation. They performed simulations on coarse grids ($10 \times 10$) and low Rayleigh numbers ($\Ra = 10^{4}, 10^{5}$) for different values of the dissipation number and concluded that Rayleigh-Bénard convection was significantly affected when the dissipation number was of order unity. 

% Fand and Bruckner \cite{fand1983} showed that inclusion of viscous dissipation leads to more accurate heat transfer correlations for the case of natural convection from horizontal cylinders.
% Viscous dissipation
% - Gebhart \cite{gebhart1962}
% - Turcotte \cite{turcotte1974}
% - Earth Mantle convection, McKenzie \cite{mckenzie1974}, Schubert \cite{schubert2004}
% - Barletta \cite{barletta2009a}

%viscous dissipation is included in compressible convection in the Earth mantle.

From an energy perspective, the viscous dissipation source term in the internal energy equation occurs as a sink in the kinetic energy equation, which cancel each other when considering the total energy equation. However, most energy analyses, especially for incompressible flow, focus on the role of the potential energy term and its split into available and background potential energy \cite{winters1995,hughes2013,gayen2013}, or on the kinetic energy budget \cite{petschel2015}. To the author's knowledge, the role of viscous dissipation in the kinetic energy equation and its numerical treatment for the internal energy equation have not been explored in detail.
 % -- being a consequence of the momentum equation -- 

In this paper, the main novelty is that we propose a discretization of the viscous dissipation function and apply it to the context of natural convection flow, where it appears as a source term in the internal energy equation. Our discretization is such that we get a correct global energy balance, on continuous, semi-discrete, and fully discrete level. First, on the continuous level, a non-dimensionalization is proposed that makes the internal and kinetic energy scaling consistent. Second, on the semi-discrete level, we propose a discrete dissipation operator, and show that it cannot be chosen freely but is \textit{implied} by the discretization of the viscous terms in the momentum equations and by the definition of the kinetic energy. This discrete dissipation operator is not only of use in the internal energy equation, but also useful beyond the context of natural convection flows, e.g.\ when estimating the dissipation of kinetic energy in turbulent flows in a numerical simulation. Third, on the fully discrete level, we propose a time integration method that preserves the total energy balance upon time marching. 
% flows We derive the local kinetic energy balance on a staggered grid,   derivation effectively shows what the form is of the discrete dissipation operator on 

% They did not provide many details of their discretization scheme and did not consider the total energy balance.

%  Our system of equations is similar to the `Extended Boussinesq approximation' considered in their work, but still includes time-dependent terms in the momentum equation. Similarly, \cite{businger2001} include the effect of viscous dissipation in atmospheric simulations. 

The paper is structured as follows. Section \ref{sec:energy_conserving_formulation} introduces the governing equations, energy balances, and new non-dimensionalization. Sections \ref{sec:energy_consistent_discretization} and \ref{sec:temporal_discretization} describe the energy-consistent spatial and temporal discretization. Section \ref{sec:results_steady} describes steady-state results of Rayleigh-Bénard convection including viscous dissipation, and section \ref{sec:results_timedependent} describes energy-conserving simulations of Rayleigh-Taylor instabilities including viscous dissipation. Section \ref{sec:results_RBC3D} shows the effect of viscous dissipation in 3D DNS of Rayleigh-Bénard convection.

\section{Energy-conserving formulation}\label{sec:energy_conserving_formulation}
\subsection{Governing equations}
The Boussinesq approximation states that density variations are small and can be ignored in all terms of the Navier-Stokes (NS) equations, except in the one pertaining to the gravity term. The NS equations describing conservation of mass and momentum then read
\begin{align}
    \nabla \cdot \vt{u} &= 0,  \label{eqn:div_free}\\
    \rho_{0} \left( \frac{\partial \vt{u}}{\partial t} + \nabla \cdot \left(\vt{u} \otimes \vt{u}\right) \right) &=  -\nabla p + \mu \nabla^2 \vt{u} + \rho \vt{g}, \label{eqn:NS_Boussinesq}
\end{align}{
where $\vt{u}(\vt{x},t)$ is the velocity field, $p(\vt{x},t)$ the pressure, $\mu$ the dynamic viscosity, $\rho(\vt{x},t)$ the density and $\rho_0$ a reference density. Without loss of generality, we consider a two-dimensional domain $\Omega$, with the gravity vector pointing in the negative $y$-direction so that $\vt{g} = - g \vt{e}_{y}$. An example of the domain as used in the Rayleigh-Bénard problem, including the boundary conditions, is given in figure \ref{fig:problem_setup}. In the results section we will also consider the Rayleigh-Taylor problem, which has adiabatic boundaries on top and bottom, instead of isothermal as in case of Rayleigh-Bénard.
% In addition, we  supplemented with the divergence-free condition
% \begin{equation}
%     \nabla \cdot \vt{u} = 0.
% \end{equation}
\begingroup
\begin{figure}[hbtp]
\fontfamily{lmss} % Latin Modern Sans Serif
%\fontfamily{phv} % helvetica
\fontsize{11pt}{11pt}\selectfont
\centering 
\def\svgwidth{0.6 \textwidth}
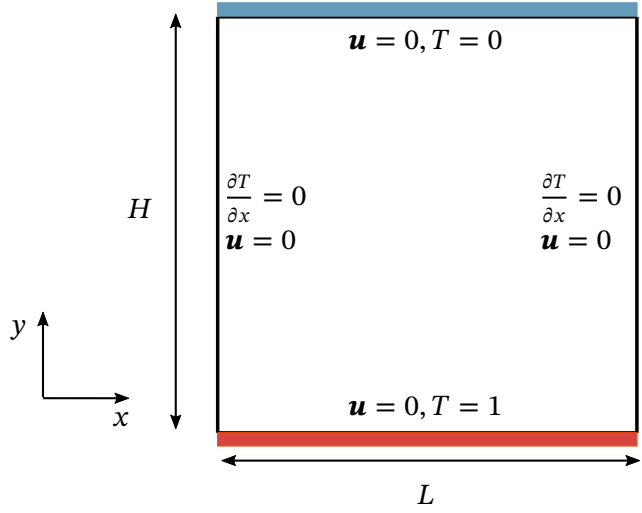 
\caption{Problem set-up for Rayleigh-Bénard convection.}
\label{fig:problem_setup}
\end{figure}
\endgroup

The density $\rho$ is assumed to vary only with temperature $T(\vt{x},t)$, according to $\rho(T) = \rho_0 - \beta \rho_0 (T - T_0)$, 
where $\beta$ is the isobaric coefficient of thermal expansion ($\beta = -\frac{1}{\rho} \left(\dd{\rho}{T} \right)_{p}$). The NS equations are then written as 
\begin{equation}\label{eqn:NS_Boussinesq_p}
    \rho_{0} \left( \frac{\partial \vt{u}}{\partial t} + \nabla \cdot \left(\vt{u} \otimes \vt{u}\right) \right) =  -\nabla p' + \mu \nabla^2 \vt{u} - \beta \rho_{0} (T - T_0) \vt{g},
\end{equation}
where $p = p' - \rho_{0} g y$ and $\nabla p = \nabla p' - \rho_{0} g \vt{e}_{y}$. 

The equation for the internal energy $e_{i}$ describes the temperature evolution according to
\begin{equation}\label{eqn:T_Boussinesq}
\dd{}{t}(\underbrace{\rho_{0} c T}_{e_{i}}) + \nabla \cdot (\vt{u} (\rho_{0} c T) ) =  \Phi +  \lambda \nabla^2 T,
\end{equation}
where $\lambda$ is the thermal conductivity and $c$ equals $c_v$ in case of an ideal gas (the specific heat at constant volume), and equals $c_p - \frac{p \beta}{\rho}$ for a real gas \cite{barletta2009}). The contribution of pressure work to the change in internal energy, $p \nabla \cdot \vt{u}$, has been discarded in equation \eqref{eqn:NS_Boussinesq_p} because of equation \eqref{eqn:div_free}.
The viscous dissipation function
\begin{equation}\label{eqn:Phi}
    \Phi := \mu \| \nabla \vt{u} \|^2 \geq 0,
\end{equation}
is the key quantity in this work, where $\| \nabla \vt{u} \|^2 = \nabla \vt{u} : \nabla \vt{u}$ (the Frobenius inner product). In 2D and Cartesian coordinates it can be written as 
\begin{equation}\label{eqn:phi_2D}
    \Phi = \mu \left[ \left( \dd{u}{x} \right)^2 + \left( \dd{u}{y} \right)^2 + \left( \dd{v}{x} \right)^2 + \left( \dd{v}{y} \right)^2 \right].
\end{equation}

\begin{remark}\label{sec:remark_dissipation_form}
The viscous dissipation expression \eqref{eqn:Phi} is only valid if the diffusive terms are written as $\nabla \cdot \vt{\tau} = \mu \nabla^2 \vt{u}$, with $\vt{\tau} = \nabla \vt{u}$. This simplified form follows from the more general stress tensor expression
\begin{equation}
    \nabla \cdot \hat{\vt{\tau}}, \qquad \hat{\vt{\tau}} = \mu (\nabla \vt{u} + (\nabla \vt{u})^T),% - \frac{2}{3} \mu \nabla \cdot \vt{u})
\end{equation}
by assuming constant $\mu$ and incompressibility, so that the identity $\nabla \cdot (\mu \nabla \vt{u})^T = \mu \nabla (\nabla \cdot \vt{u}) = 0$ holds. For the more general stress tensor $\hat{\vt{\tau}}$ we have instead
\begin{equation}\label{eqn:phi_2D_general}
    \hat{\Phi} = \hat{\vt{\tau}} : \nabla \vt{u} = \mu \left[ 2 \left(\dd{u}{x}\right)^2 + 2 \left(\dd{v}{y}\right)^2 +  \left(\dd{u}{y} + \dd{v}{x}\right)^2 \right],
\end{equation}
which is notably different from expression \eqref{eqn:phi_2D}. The dissipation function is therefore linked to the form of the stress tensor used in the momentum equations. In this work we mainly use expression \eqref{eqn:phi_2D}, but we will also explain the extension to the more general form \eqref{eqn:phi_2D_general}, see \ref{sec:discrete_kinetic_energy}, equation \eqref{eqn:Phi_hat}. For an alternative approach, see also \cite{trias2013a}.
\end{remark}

\subsection{Total energy conservation}
Conservation of kinetic energy follows by taking the dot product of equation \eqref{eqn:NS_Boussinesq_p} with $\vt{u}$:
\begin{equation}\label{eqn:kinetic_energy}
    \dd{}{t} ( \underbrace{\frac{1}{2} \rho_{0} | \vt{u} |^{2}}_{e_{k}} ) + \nabla \cdot (\frac{1}{2} \rho_{0}  |\vt{u}|^2 \vt{u}) = -\vt{u} \cdot \nabla p' +  \mu \nabla \cdot ( \vt{u} \cdot \nabla \vt{u}) - \mu  \| \nabla \vt{u} \|^2  + \beta g \rho_{0} (T - T_0) v,
\end{equation}
where $\vt{g} \cdot \vt{u} = - g v$ and we have used the identity 
\begin{equation}\label{eqn:diffusion_dissipation}
    \vt{u} \cdot \nabla^2 \vt{u} = -  \| \nabla \vt{u} \|^2 + \nabla \cdot ( \vt{u} \cdot \nabla \vt{u}).
\end{equation}

Upon adding the kinetic and internal energy equations \eqref{eqn:T_Boussinesq} and \eqref{eqn:kinetic_energy}, the viscous dissipation term cancels and we arrive at the equation for the total energy $e = \ek + \eint$:
\begin{equation}\label{eqn:total_energy}
    \dd{}{t} ( \ek + \eint ) + \nabla \cdot ( (\ek + \eint) \vt{u}) = -\nabla \cdot (p' \vt{u}) +  \mu \nabla \cdot ( \vt{u} \cdot \nabla \vt{u}) + \beta g \rho_{0} (T - T_0) v +  \lambda \nabla^2 T.
\end{equation}
All terms are in conservative (divergence) form, except the potential energy term. Upon integrating over the domain $\Omega$ and assuming no-slip conditions $\vt{u}=\vt{0}$ on all boundaries, we obtain the global balances
\begin{align}
    \frac{\rd E_{k}}{\rd t} &= - \int_{\Omega} \Phi \rd \Omega + \int_{\Omega} \beta g \rho_{0} (T - T_0) v \rd \Omega \label{eqn:kineticenergy_balance}, \\
    \frac{\rd E_{i}}{\rd t} &= \int_{\Omega} \Phi \rd \Omega + \int_{\partial \Omega} \lambda \nabla T \cdot \vt{n} \, \rd S, \label{eqn:internalenergy_balance} \\
    \frac{\rd E}{\rd t} &=\frac{\rd E_{k}}{\rd t} +\frac{\rd E_{i}}{\rd t} =  \int_{\Omega} \beta g \rho_{0} (T - T_0) v \rd \Omega + \int_{\partial \Omega} \lambda \nabla T \cdot \vt{n} \, \rd S, \label{eqn:totalenergy_balance}
\end{align}
where $E = \int_{\Omega} e \, \rd \Omega = E_{k} + E_{i}$. In case the boundary conditions are adiabatic ($\nabla T \cdot \vt{n}=0$), the last term in \eqref{eqn:totalenergy_balance} vanishes and the total energy equation expresses that the sum of internal and kinetic energy changes due to the buoyancy flux $\int_{\Omega} \beta g \rho_{0} (T - T_0) v \, \rd \Omega$ -- this case will be dealt with in the Rayleigh-Taylor set-up in section \ref{sec:results_timedependent}.

In most studies of Rayleigh-Bénard convection the dissipation function $\Phi$ is left out from the internal energy equation \eqref{eqn:T_Boussinesq}, while its corresponding counterpart in the momentum equation ($\mu \nabla^2 \vt{u}$) is still included. As a consequence, the energy lost in the kinetic energy equation is not balanced by the heat generated in the internal energy equation, so that the total energy equation features a dissipation term, which destroys the global energy balance.

\begin{remark}\label{sec:remark_potential_energy}
Equations \eqref{eqn:kineticenergy_balance} and \eqref{eqn:totalenergy_balance} feature the buoyancy flux $\int \beta g \rho_0 (T-T_0) v \rd \Omega$ (stemming from the term $\int \rho \vt{g} \cdot \vt{u} \rd \Omega$). In general compressible fluids, i.e.\ those that satisfy 
\begin{equation}\label{eqn:mass_conservation}
    \dd{\rho}{t} + \nabla \cdot (\rho \vt{u}) =0,
\end{equation}
one can show that the buoyancy flux can be written as the time derivative of the potential energy $E_{p} = \int \rho g y \rd \Omega$ (see \cite{smith2019}, section 6.4.2; \cite{kee2003}, section 3.8). In that case, one could define $\hat{E} = E_{k} + E_{i} + E_{p}$ and have a total energy conservation statement of the form \cite{tailleux2009}:
\begin{equation}\label{eqn:totalenergyconservation_withpotential}
\frac{\rd \hat{E}}{\rd t}  = \int_{\partial \Omega} \lambda \nabla T \cdot \vt{n} \, \rd S.
\end{equation}
However, in Boussinesq fluids, equation \eqref{eqn:mass_conservation} is \textit{not} satisfied; instead we have
\begin{equation}
    \dd{\rho}{t} + \nabla \cdot (\rho \vt{u}) = \dd{\rho}{t} + (\vt{u} \cdot \nabla) \rho + \underbrace{\rho \nabla \cdot \vt{u}}_{=0} = -\rho_0 \beta \left[ \dd{T}{t} + (\vt{u} \cdot \nabla) T \right].
\end{equation}
The right-hand side can be written in terms of the sum of thermal diffusion and viscous dissipation, see equation \eqref{eqn:T_Boussinesq}, and is generally nonzero. As a consequence, the time derivative of the potential energy includes not only the buoyancy flux, but also additional terms \cite{winters1995,hughes2013}. Therefore, for Boussinesq fluids additional terms appear in the right-hand side of equation \eqref{eqn:totalenergyconservation_withpotential} (independent of whether viscous dissipation is included in the internal energy equation). In this paper, the meaning `energy-consistent' thus refers to the exchange between internal and kinetic energy, and not to the total energy (kinetic + potential + internal), which is not conserved under the Boussinesq approximation.
\end{remark}

% If we further assume there are no heat sources and that the thermal conductivity $\lambda$ is constant, we get the following differential equation forms describing the energy evolution of the Boussinesq system. 
% The internal energy
% \begin{equation}\label{eqn:internal_energy_differential}
% \dd{}{t}(\underbrace{\rho_{0} c_{v} T}_{\eint) + \nabla \cdot (\vt{u} (\rho_{0} c_{v} T) ) =  \mu \| \nabla \vt{u} \|^2 +  \lambda \nabla^2 T,
% \end{equation}
% the kinetic energy
% \begin{equation}\label{eqn:kinetic_energy_differential}
%     \dd{}{t} ( \underbrace{\frac{1}{2} \rho_{0} | \vt{u} |^{2}}_{e_{k}} ) + \nabla \cdot (\frac{1}{2} \rho_{0}  |\vt{u}|^2 \vt{u}) = -\vt{u} \cdot \nabla p +  \mu \nabla \cdot ( \vt{u} \cdot \nabla \vt{u}) - \mu  \| \nabla \vt{u} \|^2  +  \rho \vt{g} \cdot \vt{u},
% \end{equation}
% and total energy

\subsection{Non-dimensionalization}
The study of the Rayleigh-Bénard convection problem is simplified by introducing dimensionless quantities. As explained in \cite{barenblatt2003}, p.\ 46, three dimensionless groups (or `similarity parameters') are needed to fully describe the problem. An important question that we address here is how the choice of non-dimensionalization changes the total energy equation. % known as the Rayleigh number, the Prandtl number and the Gebhart number.

% We return to the non-perturbation equations \eqref{eqn:NS_Boussinesq_p_app}-\eqref{eqn:T_Boussinesq_app} 
We non-dimensionalize equations \eqref{eqn:div_free}, \eqref{eqn:NS_Boussinesq} and \eqref{eqn:T_Boussinesq} by taking a reference length $H$ (cavity height), a reference temperature difference $\Delta T$ (difference between the cold and hot plates), and a reference velocity yet to be specified. From these choices we find the time scale $H/\uref$ and the pressure scale $\rho_0 \uref^2$. 
% the free fall velocity $\uref=\sqrt{\beta g \Delta T H}$ \footnote{Another common option is to take a velocity scale $\uref = \kappa/H$, with $\kappa$ defined below; this leads to different non-dimensional form.}. 
The non-dimensional quantities are thus
\begin{align}
\vt{\tilde{x}} &= \frac{\vt{x}}{H}, & \tilde{t} &= \frac{t \uref}{H}, & 
\vt{\tilde{u}} &= \frac{\vt{u}}{\uref}, & \tilde{T} &= \frac{T - T_{0}}{\Delta T}, & \tilde{p}' &= \frac{p'}{\rho_0 \uref^2},
\end{align}
% Substitution into the (non-perturbation) equations \eqref{eqn:NS_Boussinesq_p_app} gives
% \begin{equation}\label{eqn:NS_Boussinesq_nondim}
%     \frac{\rho_0 \uref^2}{H} \left( \frac{\partial \vt{\tilde{u}}}{\partial \tilde{t}} + \tilde{\nabla} \cdot \left(\vt{\tilde{u}} \otimes \vt{\tilde{u}}\right) \right) =  - \frac{\rho_0 \uref^2}{H} \nabla \tilde{p}' + \mu \frac{\uref}{H^2} \tilde{\nabla}^2 \vt{\tilde{u}} - \beta \rho_{0} \Delta T \tilde{T} \vt{g},
% \end{equation}
and the non-dimensional equations read
\begin{align}
    \tilde{\nabla} \cdot \vt{\tilde{u}} &= 0, \label{eqn:mass_nondim}\\
     \frac{\partial \vt{\tilde{u}}}{\partial \tilde{t}} + \tilde{\nabla} \cdot \left(\vt{\tilde{u}} \otimes \vt{\tilde{u}}\right) &=  - \tilde{\nabla} \tilde{p}' +  \frac{\mu }{\rho_{0} \uref H} \tilde{\nabla}^2 \vt{\tilde{u}} + \frac{\beta g  \Delta T H}{\uref^2} \tilde{T}  \vt{e}_{y}, \label{eqn:NS_Boussinesq_nondim2}\\
      \dd{\tilde{T}}{\tilde{t}} + \tilde{\nabla} \cdot (\vt{\tilde{u}} \tilde{T} ) &= \frac{\nu \uref}{c H \Delta T} \tilde{\Phi} +  \frac{\kappa}{\uref H}  \tilde{\nabla}^2 \tilde{T},
\end{align}
where $\nu = \mu/\rho_0$ and $\kappa = \lambda / (\rho_0 c)$. The two latter equations are re-written by introducing the parameters $\alpha_{i}$, $i=1 \ldots 4$, as
\begin{align}
     \frac{\partial \vt{\tilde{u}}}{\partial \tilde{t}} + \tilde{\nabla} \cdot \left(\vt{\tilde{u}} \otimes \vt{\tilde{u}}\right) &=  - \tilde{\nabla} \tilde{p}' +  \alpha_{1}  \tilde{\nabla}^2 \vt{\tilde{u}} + \alpha_{2}  \tilde{T}  \vt{e}_{y}, \label{eqn:NS_boussinesq_nondim3} \\
     \dd{\tilde{T}}{\tilde{t}} + \tilde{\nabla} \cdot (\vt{\tilde{u}} \tilde{T} )  &= \alpha_{3} \tilde{\Phi} +  \alpha_{4}   \tilde{\nabla}^2 \tilde{T}.\label{eqn:T_Boussinesq_nondim3}
\end{align}

The $\alpha_{i}$'s can be expressed in terms of three dimensionless numbers, being the Rayleigh number $\Ra$, the Prandtl number $\Pr$ and the Gebhart number $\Ge$ (also known as the dissipation number \cite{schubert2001}):
\begin{align}
 \Ra &= \frac{\beta g \Delta T H^3}{\nu \kappa}, \\   
 \Pra &= \frac{\nu}{\kappa}, \\
 \Ge &= \frac{\beta g H}{c}. 
\end{align}
Alternatively, one can employ the Grashof number $\text{Gr} = \Ra / \Pr$ \cite{barenblatt2003}. In table \ref{tab:nondim_forms} we present three different options for $\uref$ with the corresponding values of $\alpha$. Choices I and II are common in literature, see for example \cite{vanderpoel2013} for choice I and \cite{barletta2009a,siggia1994,hepworth2014} for choice II. Other choices are also possible, e.g.\ $\uref = \beta g \Delta T H^2 / \nu$ \cite{mckenzie1974}, but this choice does not lead to a `clean' expression in terms of the dimensionless numbers defined above. To our best knowledge, choice III is new and inspired by the form of the total energy equation, as we will explain below. %in section \ref{sec:continuous_energy_balance}.

\begin{table}[h]
    \centering
    \begin{tabular}{c c c c c c c}
        \toprule
          & $\uref$  & $\alpha_1=\frac{\nu}{\uref H}$ & $\alpha_2=\frac{\beta g \Delta T H}{\uref^2}$ &$\alpha_3=\frac{\nu \uref}{c \Delta T H}$ & $\alpha_4= \frac{\kappa}{\uref H}$ & $\gamma = \frac{\alpha_{1}}{\alpha_{3}}$  \\
         \midrule
         I & $\sqrt{\beta g \Delta T H}$  & $\sqrt{\frac{\Pra}{\Ra}}$ & 1 & $\Ge \sqrt{\frac{\Pra}{\Ra}}$ &  $\frac{1}{\sqrt{\Pra \Ra}}$ & $\frac{1}{\Ge}$\\
         II & $\frac{\kappa}{H}$  & $\Pra$ & $\Pra \Ra$ &  $\frac{\Ge}{\Ra}$ & 1 & $\frac{\Pra \Ra}{\Ge}$ \\
         III & $\sqrt{c \Delta T}$  & $\sqrt{\frac{\Pra \Ge}{\Ra}}$ & $\Ge$ & $\sqrt{\frac{\Pra \Ge}{\Ra}}$ & $\sqrt{\frac{\Ge}{\Pra \Ra}}$  & 1 \\
         \bottomrule
    \end{tabular}
    \caption{Different non-dimensional forms resulting from different choices of $\uref$.}
    \label{tab:nondim_forms}
\end{table}

It is important to realize that the time scales and the velocity fields corresponding to numerical simulations with choices I, II and III are different. The time scales are related as $\frac{\tilde{t}_{I}}{u_{\text{ref},{I}}} = \frac{\tilde{t}_{II}}{u_{\text{ref},{II}}} = \frac{\tilde{t}_{III}}{u_{\text{ref},{III}}}$, so $\tilde{t}_{III} = \tilde{t}_{I} / \sqrt{\Ge}$ and 
    $\tilde{t}_{III} =\tilde{t}_{II} \sqrt{\Ra \Pr} / \sqrt{\Ge}$.
% \begin{equation}
%     t = \frac{\tilde{t}_{I} H}{u_{\text{ref},{I}}} = \frac{\tilde{t}_{II} H}{u_{\text{ref},{II}}} = \frac{\tilde{t}_{III} H}{u_{\text{ref},{III}}},
% \end{equation}
% so that a simulation performed with our proposed choice III will have its time scale related to choice I according to $\tilde{t}_{III} = \tilde{t}_{I} / \sqrt{\Ge}$, and related to choice II according to $\tilde{t}_{III}=\tilde{t}_{II} \sqrt{\Ra \Pr/ \Ge}$.
The velocity fields are related as $\vt{\tilde{u}}_{I} u_{\text{ref},{I}} = \vt{\tilde{u}}_{II} u_{\text{ref},{II}} = \vt{\tilde{u}}_{III} u_{\text{ref},{III}}$, so that 
% \begin{align}
    $\vt{\tilde{u}}_{III} = \vt{\tilde{u}}_{I} \sqrt{\Ge}$, and 
    $\vt{\tilde{u}}_{III} = \vt{\tilde{u}}_{II} \sqrt{\Ge} / \sqrt{\Ra \Pr}$.
% \end{align}
% \begin{equation}
%     \vt{u} = \vt{\tilde{u}}_{I} u_{\text{ref},{I}} = \vt{\tilde{u}}_{II} u_{\text{ref},{II}} = \vt{\tilde{u}}_{III} u_{\text{ref},{III}}.
% \end{equation}
 On the other hand, the temperature fields corresponding to each choice are equivalent, and consequently the Nusselt numbers are the same.

% \subsection{Energy balances}\label{sec:continuous_energy_balance}
To obtain the non-dimensional form of the total energy equation we take the dot product of \eqref{eqn:NS_boussinesq_nondim3} with $\vt{\tilde{u}}$ and add the internal energy equation \eqref{eqn:T_Boussinesq_nondim3}. In order for the dissipation function of the kinetic energy equation to cancel with the internal energy equation, we require $\alpha_{1} = \alpha_{3}$. This requirement is satisfied by $\uref = \sqrt{c \Delta T}$, i.e.\ our proposed choice III in table \ref{tab:nondim_forms}. For the other choices (I and II), a weighting of the kinetic and internal energy equations is needed in order to cancel the dissipation function in the non-dimensional total energy equation. The weighting factor depends on the definition of the non-dimensional total energy. First define the dimensionless kinetic and internal energy as
\begin{align}
    \tilde{e}_{k} &:= \frac{\ek}{\rho_{0} \uref^{2} } = \frac{\frac{1}{2} \rho_{0} |\vt{u}|^{2}}{\rho_{0} \uref^{2}} = \frac{\frac{1}{2} \rho_{0} \uref^2 |\vt{\tilde{u}}|^{2}}{\rho_{0} \uref^{2}} = \frac{1}{2}|\vt{\tilde{u}}|^{2}, \\
    %\ek &= \frac{1}{2} \rho_{0} |\vt{u}|^{2} = \rho_{0} \uref^{2}  \frac{1}{2} |\vt{\tilde{u}}|^2 =: \rho_{0} \uref^{2} \tilde{e}_{k}, \\
    \tilde{e}_{i} &:= \frac{\eint}{\rho_{0} c \Delta T} = \frac{\rho_{0} c T}{\rho_{0} c \Delta T} = \frac{\rho_{0} c \Delta T (\tilde{T} + T_0/\Delta T)}{\rho_{0} c \Delta T} = (\tilde{T} + T_0/\Delta T), %\eint &=  \rho_{0} c T = \rho_{0} c \Delta T (\tilde{T} + T_0/\Delta T) =: \rho_{0} c \Delta T \tilde{e}_{i},
\end{align}
so that
\begin{equation}
    e = \ek + \eint = \rho_{0} \uref^{2} \tilde{e}_{k} + \rho_{0} c \Delta T \tilde{e}_{i} =  \rho_{0} \uref^{2} \left( \tilde{e}_{k} + \frac{ c \Delta T}{\uref^2} \tilde{e}_{i} \right).
\end{equation}
By \textit{choosing} the non-dimensional total energy as $\tilde{e} = e/ \rho_{0} \uref^2$, we obtain 
\begin{equation}
    \tilde{e} = \tilde{e}_{k} + \frac{ c \Delta T}{\uref^2} \tilde{e}_{i} = \tilde{e}_{k} + \frac{\alpha_{1}}{\alpha_{3}} \tilde{e}_{i} = \tilde{e}_{k} + \gamma \tilde{e}_{i}.
\end{equation}
Here $\gamma=\frac{\alpha_{1}}{\alpha_{3}}$ is the weighting factor, which is reported in table \ref{tab:nondim_forms} for different choices of $\uref$. The global energy balances in non-dimensional form read
\begin{align}
    \frac{\rd \tilde{E}_{k}}{\rd \tilde{t}} &= - \frac{\alpha_{1}}{\Lambda } \int_{\tilde{\Omega}} \tilde{\Phi} \, \rd \tilde{\Omega} + \frac{\alpha_{2}}{\Lambda } \int_{\tilde{\Omega}} \tilde{T} \tilde{v} \, \rd \tilde{\Omega}, \label{eqn:kineticenergy_balance_nondim} \\
    \frac{\rd \tilde{E}_{i}}{\rd \tilde{t}} &= \frac{\alpha_{3}}{\Lambda } \int_{\tilde{\Omega}} \tilde{\Phi} \, \rd \tilde{\Omega} + \frac{\alpha_{4}}{\Lambda} \int_{\partial \tilde{\Omega}} \tilde{\nabla} \tilde{T} \cdot \vt{n} \, \rd \tilde{S}, \label{eqn:internalenergy_balance_nondim} \\
    \frac{\rd \tilde{E}}{\rd \tilde{t}} &= \frac{\rd \tilde{E}_{k}}{\rd t} + \gamma \frac{\rd \tilde{E}_{i}}{\rd t} = \frac{\alpha_{2}}{\Lambda } \int_{\tilde{\Omega}} \tilde{T} \tilde{v} \, \rd \tilde{\Omega} + \frac{\gamma \alpha_{4}}{\Lambda} \int_{\partial \tilde{\Omega}} \tilde{\nabla} \tilde{T} \cdot \vt{n} \, \rd \tilde{S}, \label{eqn:totalenergy_balance_nondim}
\end{align}
where we define $\tilde{E} = \frac{1}{\Lambda} \int_{\tilde{\Omega}} \tilde{e} \, \rd \tilde{\Omega}$, and $\Lambda = L/H$ is the aspect ratio of the box.

The choice for a particular reference velocity typically depends on the problem at hand. Choices I and II have the advantage that in case of $\Ge =0$ (most commonly investigated in literature), one obtains $\alpha_{3}=0$ and the dissipation terms simply drops from the internal energy equation. However, when $\Ge$ is small but nonzero, the weight factor $\gamma$ becomes very large for choices I and II. Choice III does not suffer from this issue, because $\gamma=1$ independent of $\Ge$. However, choice III has the disadvantage that it does not work in the case $\Ge=0$, since it leads to $\alpha_{i}=0$ for all $i$. In summary: for $\Ge=0$, choices I and II are preferred; for small but nonzero $\Ge$, choice III is preferred; in other cases, all choices are fine.
% In addition, the third choice is elegant as no dimensionless groups appear in the definition of the dimensionless total energy. The `disadvantage' of this choice is that $\Ge$ appears in all $\alpha$'s, so that the case $\Ge = 0$ is not well-covered. Choices I and II lead to the situation that for $\Ge=0$ the dissipation term simply disappears from the temperature equation.

\subsection{Effect of viscous dissipation on Nusselt number and thermal dissipation}
A main quantity of interest in natural convection flows is the Nusselt number $\Nu$ and we will investigate how it changes upon including viscous dissipation in the internal energy equation. First, define the average of the sum of convective and conductive fluxes through a horizontal plane $y=y'$ by
\begin{equation}
    \overline{F}(y') := \frac{1}{L} \int_{0}^{L} \left( \rho_{0} c T v - \lambda \dd{T}{y} \right)_{(x,y')} \, \rd x.
\end{equation}
%Non-dimensionalizing in the same way as the internal energy equation leads to the following expression, which 
Then, the Nusselt number based on $\overline{F}$ follows as \cite{grossmann2000}:
\begin{equation}\label{eqn:Nusselt_general}
    \Nu(\tilde{y}') := \frac{\overline{F}(y')}{\lambda \Delta T/H} = \frac{1}{\Lambda} \int_{0}^{\Lambda} \left( \frac{1}{\alpha_{4}} \tilde{T} \tilde{v} - \dd{\tilde{T}}{\tilde{y}} \right)_{(\tilde{x},\tilde{y}')} \, \rd \tilde{x}.
\end{equation}
For steady state or statistically steady state (using a suitable average), and in the absence of viscous dissipation, it is straightforward to show from the internal energy equation that $\Nu(\tilde{y}) = \Nu(\tilde{y}=0) = \Nu$, which is a constant, independent of $\tilde{y}'$ \cite{siggia1994,hepworth2014}. However, upon including viscous dissipation, this relation no longer holds true and instead the steady internal energy equation yields
\begin{equation}\label{eqn:Nusselt_epsU_y}
    \alpha_{4} (\Nu(\tilde{y}') - \Nu(0)) = \alpha_{3} \epsilon_{U}(\tilde{y}'),
\end{equation}
where the integrated dissipation function is given by
\begin{equation}\label{eqn:epsU_phi}
    \epsilon_{U}(\tilde{y}') := \frac{1}{\Lambda} \int_{0}^{\tilde{y}'} \int_{0}^{\Lambda} \tilde{\Phi} \, \rd \tilde{x} \rd \tilde{y}.
\end{equation}
Equation \eqref{eqn:Nusselt_epsU_y} is an important relation which shows that (taking $\tilde{y}'=1$)
\begin{equation}\label{eqn:Nusselt_epsU}
\alpha_{4}(\Nu(1) - \Nu(0)) = \alpha_{3} \epsilon_{U}(1),
\end{equation}
so \textit{the Nusselt number of the upper plate is always larger than or equal to the Nusselt number of the lower plate}. 
%(with equality only for the trivial solution $\vt{\tilde{u}}=0$)

A second relation between Nusselt number and viscous dissipation can be obtained from the global kinetic energy balance, equation \eqref{eqn:kineticenergy_balance_nondim}. 
% In non-dimensional form we obtain
% \begin{equation}\label{eqn:kineticenergy_global_nondim}
%     \frac{1}{\Lambda }\frac{\rd }{\rd t} \int_{\tilde{\Omega}} \frac{1}{2} |\vt{\tilde{u}}|^2 \, \rd \tilde{\Omega} = - \frac{\alpha_{1}}{\Lambda } \int_{\tilde{\Omega}} \tilde{\Phi} \, \rd \tilde{\Omega} + \frac{\alpha_{2}}{\Lambda } \int_{\tilde{\Omega}} \tilde{T} \tilde{v} \, \rd \tilde{\Omega}.
% \end{equation}
The second term in the right-hand side of equation \eqref{eqn:kineticenergy_balance_nondim} can be rewritten with equation \eqref{eqn:Nusselt_epsU_y}, following the analysis in \cite{hepworth2014}:
\begin{equation}
    \begin{split}
        \frac{\alpha_{2}}{\Lambda } \int_{\tilde{\Omega}} \tilde{T} \tilde{v} \, \rd \tilde{\Omega} &= \frac{\alpha_{2}}{\Lambda } \int_{0}^{1} \int_{0}^{\Lambda} \tilde{T} \tilde{v} \, \rd \tilde{x} \rd \tilde{y}  
        = \alpha_{2} \alpha_{4} \int_{0}^{1} \Nu(\tilde{y}) \, \rd \tilde{y} + \frac{\alpha_{2} \alpha_{4}}{\Lambda} \int_{0}^{\Lambda} \int_{0}^{1} \dd{\tilde{T}}{\tilde{y}} \, \rd \tilde{y} \rd \tilde{x} \\
        &= \alpha_{2} \alpha_{4} \Nu(0) + \alpha_{2} \alpha_{3} \int_{0}^{1} \epsilon_{U} (\tilde{y}) \, \rd \tilde{y} +  \frac{\alpha_{2} \alpha_{4}}{\Lambda} \int_{0}^{\Lambda} (\tilde{T}(\tilde{x},\tilde{y}=1) - \tilde{T}(\tilde{x},\tilde{y}=0)) \rd \tilde{x} \\
        &= \alpha_{2} \alpha_{4} (\Nu(0) - 1) + \alpha_{2} \alpha_{3} \int_{0}^{1} \epsilon_{U} (\tilde{y}) \, \rd \tilde{y}.
    \end{split}
\end{equation}
For (statistically) steady flow, this term equals the first term in the right-hand side of equation \eqref{eqn:kineticenergy_balance_nondim}, yielding the second relation between the Nusselt number and the viscous dissipation $\epsilon_{U}$
\begin{equation}\label{eqn:Nusselt_kinenergy}
     \alpha_{2} \alpha_{4} (\Nu(0) - 1) = \alpha_{1} \epsilon_{U}(1) - \alpha_{2} \alpha_{3}\int_{0}^{1} \epsilon_{U} (\tilde{y}) \, \rd \tilde{y}.    
\end{equation}
We recognize the well-known equation $\alpha_{2} \alpha_{4} (\Nu(0) - 1) = \alpha_{1} \epsilon_{U}(1)$, see e.g.\ \cite{siggia1994}, but with the additional negative term $- \alpha_{2} \alpha_{3}\int_{0}^{1} \epsilon_{U} (\tilde{y}) \, \rd \tilde{y}$. %The deviation of the Nusselt number from unity is therefore decreased by the presence of viscous dissipation in the internal energy equation, as compared to the case without viscous dissipation in the internal energy equation.
%of \eqref{eqn:Nusselt_kinenergy} is that viscous dissipation fromThis term arises from viscous dissipation in the internal energy equation in addition to the viscous dissipation term arising from the kinetic energy equation. %which features the viscous dissipation term because the kinetic energy equation balance is considered

% with $\epsilon_{U} := \frac{1}{\Lambda} \int_{\tilde{\Omega}} \tilde{\Phi} \, \rd \tilde{\Omega}$.

Lastly, we link the thermal dissipation $\epsilon_{T}$ to the Nusselt number and the viscous dissipation function. The non-dimensional internal energy equation, equation \eqref{eqn:T_Boussinesq_nondim3}, is multiplied by $\tilde{T}$, and after integrating by parts, using the skew-symmetry of the convective operator, and employing the boundary condition $\tilde{T}(\tilde{y}=1)=0$, one obtains
\begin{equation}\label{eqn:thermal_dissipation}
    \frac{1}{\Lambda }\frac{\rd }{\rd t} \int_{\tilde{\Omega}} \frac{1}{2} \tilde{T}^2 \, \rd \tilde{\Omega} = \frac{\alpha_{3}}{\Lambda} \int_{\tilde{\Omega}} \tilde{T} \tilde{\Phi} \, \rd \tilde{\Omega} - \frac{\alpha_{4}}{\Lambda} \int_{0}^{\Lambda} \left(\tilde{T} \dd{\tilde{T}}{\tilde{y}} \right)_{\tilde{y}=0} \, \rd \tilde{x} - \frac{\alpha_{4}}{\Lambda} \int_{\tilde{\Omega}}  \| \tilde{\nabla} \tilde{T} \|^2 \, \rd \tilde{\Omega}.
\end{equation}
With the boundary condition $\tilde{T}(\tilde{y}=0)=1$, and the assumption of (statistically) steady flow, this relation is further simplified to
\begin{equation}\label{eqn:Nusselt_thermal_diss}
    \alpha_{4} \Nu(0) = \alpha_{4} \epsilon_{T} - \frac{\alpha_{3}}{\Lambda} \int_{\tilde{\Omega}} \tilde{T} \tilde{\Phi} \, \rd \tilde{\Omega},
\end{equation}
where 
\begin{equation}\label{eqn:thermal_dissipation_def}
    \epsilon_{T} := \frac{1}{\Lambda} \int_{\tilde{\Omega}} \| \tilde{\nabla} \tilde{T} \|^2 \, \rd \tilde{\Omega}.
\end{equation}
Since $\tilde{T}\geq 0$, $\tilde{\Phi}\geq 0$, we conclude that \textit{viscous dissipation lowers the Nusselt number of the lower plate}. In absence of viscous dissipation in the internal energy equation, one obtains the familiar relation $\Nu =  \epsilon_{T}$. In combination with equation \eqref{eqn:Nusselt_epsU}, we obtain for the Nusselt number of the upper plate:
\begin{equation}
\alpha_{4} \Nu(1) = \alpha_{4} \epsilon_{T} + \frac{\alpha_{3}}{\Lambda} \int (1-\tilde{T}) \tilde{\Phi} \, \rd \tilde{\Omega}.
\end{equation}
Assuming that the temperature satisfies $0\leq \tilde{T} \leq 1$, we find that \textit{viscous dissipation increases the Nusselt number of the upper plate}. In other words, the thermal dissipation lies in between the two Nusselt numbers:
\begin{equation}
    \Nu(0) \leq \epsilon_{T} \leq \Nu(1).
\end{equation}
The three relations \eqref{eqn:Nusselt_epsU}, \eqref{eqn:Nusselt_kinenergy} and \eqref{eqn:Nusselt_thermal_diss} are summarized in table \ref{tab:Nusselt_relations} and will be confirmed in the numerical experiments in section \ref{sec:results_steady}.
% \begin{equation}
%     0 \leq \frac{1}{\Lambda} \int_{\tilde{\Omega}} \tilde{T} \tilde{\Phi} \, \rd \tilde{\Omega} \leq \epsilon_{U},
% \end{equation}
% so that the Nusselt number at the lower plate is bounded by 
% \begin{equation}
%     \alpha_{4} \epsilon_{T} - \alpha_{3} \epsilon_{U} \leq \Nu(0) \leq \alpha_{4} \epsilon_{T}.
% \end{equation}
\begin{table}[h!]
    \centering
    \begin{tabular}{c c c c}
    \toprule
        origin &  without viscous dissipation & with viscous dissipation  \\
        \midrule
      internal   & $\Nu(1) = \Nu(0)$ & $\alpha_{4}(\Nu(1) - \Nu(0)) = \alpha_{3} \epsilon_{U}(1)$ \\
      kinetic & $\alpha_{2} \alpha_{4} (\Nu(0) - 1) = \alpha_{1} \epsilon_{U}(1)$ &  $\alpha_{2} \alpha_{4} (\Nu(0) - 1) = \alpha_{1} \epsilon_{U}(1) - \alpha_{2} \alpha_{3}\int_{0}^{1} \epsilon_{U} (\tilde{y}) \, \rd \tilde{y}$ \\
      internal energy $\times T$ & $ \Nu(0) = \epsilon_{T}$ &  $\alpha_{4} \Nu(0) = \alpha_{4} \epsilon_{T} - \frac{\alpha_{3}}{\Lambda} \int_{\tilde{\Omega}} \tilde{T} \tilde{\Phi} \, \rd \tilde{\Omega}$ \\
      \bottomrule
    \end{tabular}
    \caption{Steady-state Nusselt number relations, with and without viscous dissipation.}
    \label{tab:Nusselt_relations}
\end{table}

\section{Energy-consistent spatial discretization}\label{sec:energy_consistent_discretization}
\subsection{Mass, momentum and kinetic energy equation}
To discretize the non-dimensional mass and momentum equations \eqref{eqn:mass_nondim} and \eqref{eqn:NS_boussinesq_nondim3}, we use the staggered-grid energy-conserving finite volume method described in \cite{sanderse2013a}, extended by including the buoyancy term in the momentum equations. This leads to the following semi-discrete equations:
\begin{align} 
    M V_{h}(t) &= 0, \label{eqn:mass_semidiscrete} \\
    \Omega_{V} \frac{\rd V_{h}(t)}{\rd t} &= - C_{V} (V_{h}(t)) - G p_{h}(t) + \alpha_{1} D_{V} V_{h}(t) + \alpha_{2} (A  T_{h}(t) + y_{T}). \label{eqn:mom_semidiscrete}
\end{align}
Here, $V_{h} \in \mathbb{R}^{N_{V}}$ are the velocity unknowns, $p_{h} \in \mathbb{R}^{N_{p}}$ the pressure unknowns, and $T_{h} \in \mathbb{R}^{N_{p}}$ the temperature unknowns; see figure \ref{fig:staggered} for their positioning. $M \in \mathbb{R}^{N_p \times N_V}$ is the discretized divergence operator, $ G = - M^T \in \mathbb{R}^{N_V \times N_p}$ the discretized gradient operator, $\Omega_{V} \in \mathbb{R}^{N_{V} \times N_{V}}$ a matrix with the `velocity' finite volume sizes on its diagonal, and $C_{V}$ and $D_{V}$ constitute central difference approximations of the convective and diffusive terms. $A$ is a matrix that averages the temperature from the center of the `temperature' finite volumes to center of the `velocity' finite volumes, and the vector $y_{T}$ incorporates the nonzero boundary condition for the temperature at the lower plate.

The energy-conserving nature of our finite volume method is crucial in deriving an energy-consistent discretization of viscous dissipation. The energy-conserving property means that, in absence of boundary contributions, the discretized convective and pressure gradient operators do not contribute to the kinetic energy balance: $V_{h}^T C_{V} (V_{h}) = 0$ and $V_{h}^T G p_{h} = 0$, just like in the continuous case. This is achieved by using a skew-symmetric convection operator and the compatibility between $M$ and $G$ via $G=-M^{T}$. The discrete kinetic energy balance then reads:
\begin{equation}\label{eqn:kineticenergy_semidiscrete}
    \frac{\rd E_{k,h}}{\rd t} = - \alpha_{1} \epsilon_{U,h} + \alpha_{2} V_{h}^T ( A T_{h} + y_{T}),
\end{equation}
where $E_{k,h} = \frac{1}{2} V_{h}^{T} \Omega_{V} V_{h}$. The global viscous dissipation (i.e.\ summed over the entire domain) is given by $\epsilon_{U,h} = \| Q V_{h}\|_2^2 >0$, where $Q$ stems from decomposing the symmetric negative-definite diffusive operator as $ D_{V} = -Q^T Q$. Equation \eqref{eqn:kineticenergy_semidiscrete} is the semi-discrete counterpart of equation \eqref{eqn:kineticenergy_balance_nondim}. % (see also \eqref{eqn:kineticenergy_balance}).

\begingroup
\begin{figure}[hbtp]
\fontfamily{lmss} % Latin Modern Sans Serif
%\fontfamily{phv} % helvetica
\fontsize{11pt}{11pt}\selectfont
\centering 
\def\svgwidth{0.45 \textwidth}
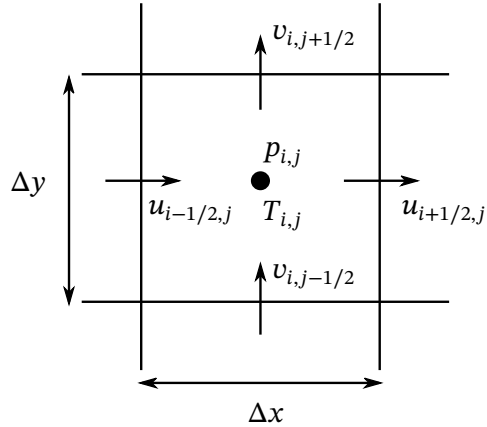 
\caption{Staggered grid with positioning of unknowns around a pressure volume.}
\label{fig:staggered}
\end{figure}
\endgroup

\subsection{Proposed viscous dissipation function}
Given a discretization that satisfies a discrete kinetic energy balance, the key step is to design a discretization scheme of the internal energy equation \eqref{eqn:T_Boussinesq_nondim3} which is such that discrete versions of the global balances \eqref{eqn:internalenergy_balance} and \eqref{eqn:totalenergy_balance} are obtained. In particular, the viscous dissipation in the internal energy equation should cancel the viscous dissipation term in the kinetic energy equation, where the latter is fully determined by \textit{the choice of the diffusion operator and the expression for the local kinetic energy}.
% This requires a discretization of the \textit{local} dissipation function $\Phi_{h} \in \mathbb{R}^{N_p}$ that is consistent with the discretization of the diffusion operator $D_{V}$
% , i.e.\ $\Phi_{h}$ should satisfy
% \begin{equation}
%     \alpha_{3} 1^{T} \Omega_{p} \Phi_{h} = \alpha_{1} \epsilon_{U,h}.
% \end{equation}
% while in the global kinetic energy equation only the global quantity $\Phi$ is needed. 
The choice for the diffusion operator (second-order central differencing) is straightforward. The choice for the expression of the local kinetic energy on a staggered grid is however not obvious. We propose the following definition:
\begin{equation}\label{eqn:local_KE}
    k_{i,j} := \frac{1}{4} u_{i+1/2,j}^2 + \frac{1}{4} u_{i-1/2,j}^2 + \frac{1}{4} v_{i,j+1/2}^2 + \frac{1}{4} v_{i,j-1/2}^2.
\end{equation}
This choice gives a local kinetic energy equation that is consistent with the continuous equations, as is detailed in \ref{sec:discrete_kinetic_energy}, and consistent with the global energy definition.

% The expression for $\Phi_{h}$ that satisfies this equation can be derived by considering the local kinetic energy equation. This is  
The expression for $\Phi_{h}$ then follows from differentiating the expression for $k_{ij}$ in time, substituting the momentum equations, and rewriting the terms involving the diffusive operator (see \ref{sec:discrete_kinetic_energy}). The implied dissipation then follows by constructing a discrete version of \eqref{eqn:diffusion_dissipation}. As example, we construct the discrete version of $u \frac{\partial^2 u}{\partial x^2} =  - \left( \frac{\partial u}{\partial x}\right)^2 + \frac{\partial}{\partial x} \left( u \frac{\partial u}{\partial x}\right)$, being
\begin{multline}\label{eqn:discrete_local_dissipation_main}
    \frac{u_{i+1/2,j}}{\Delta x} \left( \frac{u_{i+3/2,j} - u_{i+1/2,j}}{\Delta x} - \frac{u_{i+1/2,j} - u_{i-1/2,j}}{\Delta x}  \right) 
    =   - \frac{1}{2} \left( \frac{u_{i+3/2,j} - u_{i+1/2,j}}{\Delta x} \right)^2 - \frac{1}{2} \left( \frac{u_{i+1/2,j} - u_{i-1/2,j}}{\Delta x} \right)^2\\
    + \frac{1}{\Delta x} \left( \frac{1}{2}(u_{i+3/2,j} + u_{i+1/2,j})\frac{u_{i+3/2,j} - u_{i+1/2,j}}{\Delta x} - \frac{1}{2}(u_{i+1/2,j} + u_{i-1/2,j})\frac{u_{i+1/2,j} - u_{i-1/2,j}}{\Delta x} \right) .
\end{multline}
The first two terms on the right-hand side contribute to the viscous dissipation function. Repeating this process for the other components ($u \frac{\partial^2 u}{\partial y^2}$, $v \frac{\partial^2 v}{\partial x^2}$, $v \frac{\partial^2 v}{\partial y^2}$), as outlined in \ref{sec:discrete_kinetic_energy_dissipation}, yields the following novel expression for the local dissipation function:
\begin{equation}\label{eqn:local_dissipation_function}
    \boxed{\Phi_{i,j} = \frac{1}{2} \Phi^{u}_{i+1/2,j} + \frac{1}{2} \Phi^{u}_{i-1/2,j} + \frac{1}{2} \Phi^{v}_{i,j+1/2} + \frac{1}{2} \Phi^{v}_{i,j-1/2}},
\end{equation}
where
\begin{multline}
    \Phi^{u}_{i+1/2,j} = - \frac{1}{2}\left( \frac{u_{i+3/2,j} - u_{i+1/2,j}}{\Delta x} \right)^2 -  \frac{1}{2} \left( \frac{u_{i+1/2,j} - u_{i-1/2,j}}{\Delta x} \right)^2  - \frac{1}{2}\left( \frac{u_{i+1/2,j+1} - u_{i+1/2,j}}{\Delta y} \right)^2 - \frac{1}{2} \left( \frac{u_{i+1/2,j} - u_{i+1/2,j-1}}{\Delta y} \right)^2,      
\end{multline}
\begin{multline}
    \Phi^{v}_{i,j+1/2} = -  \frac{1}{2} \left( \frac{v_{i+1,j+1/2} - v_{i,j+1/2}}{\Delta x} \right)^2  - \frac{1}{2} \left( \frac{v_{i+1,j-1/2} - v_{i,j-1/2}}{\Delta x} \right)^2  - \frac{1}{2}\left( \frac{v_{i,j+3/2} - v_{i,j+1/2}}{\Delta y} \right)^2 -  \frac{1}{2} \left( \frac{v_{i,j+1/2} - v_{i,j-1/2}}{\Delta y} \right)^2. 
\end{multline}

% \begin{multline}\label{eqn:local_dissipation_function}
%     \Phi_{i,j} = - \frac{1}{4} \left( \frac{u_{i+3/2,j} - u_{i+1/2,j}}{\Delta x} \right)^2 - \frac{1}{2} \left( \frac{u_{i+1/2,j} - u_{i-1/2,j}}{\Delta x} \right)^2 -  \frac{1}{4} \left( \frac{u_{i-1/2,j} - u_{i-3/2,j}}{\Delta x} \right)^2 \\
%     - \frac{1}{4} \left( \frac{u_{i-1/2,j+1} - u_{i-1/2,j}}{\Delta y} \right)^2 - \frac{1}{4} \left( \frac{u_{i-1/2,j} - u_{i-1/2,j-1}}{\Delta y} \right)^2     - \frac{1}{4} \left( \frac{u_{i+1/2,j+1} - u_{i+1/2,j}}{\Delta y} \right)^2 - \frac{1}{4} \left( \frac{u_{i+1/2,j} - u_{i+1/2,j-1}}{\Delta y} \right)^2 \\
%     - \frac{1}{4} \left( \frac{v_{i+1,j+1/2} - v_{i,j+1/2}}{\Delta x} \right)^2   - \frac{1}{4} \left( \frac{v_{i,j+1/2} - v_{i-1,j+1/2}}{\Delta x} \right)^2     - \frac{1}{4} \left( \frac{v_{i+1,j-1/2} - v_{i,j-1/2}}{\Delta x} \right)^2   - \frac{1}{4} \left( \frac{v_{i,j-1/2} - v_{i-1,j-1/2}}{\Delta x} \right)^2 \\
%     - \frac{1}{4} \left( \frac{v_{i,j+3/2} - v_{i,j+1/2}}{\Delta y} \right)^2 -  \frac{1}{2} \left( \frac{v_{i,j+1/2} - v_{i,j-1/2}}{\Delta y} \right)^2 -  \frac{1}{4} \left( \frac{v_{i,j-1/2} - v_{i,j-3/2}}{\Delta y} \right)^2 \nonumber
% \end{multline}
At boundaries, an adaptation of $\Phi_{h}$ is required in order to have a discrete equivalent of equation \eqref{eqn:diffusion_dissipation}. This is detailed in equation \eqref{eqn:Phi_BC}.

% The temperature is, like the pressure, defined at the center of a finite volume. As a consequence, the gravity term $T \vt{e}_{y}$ in the vertical momentum equation requires interpolation, and we take simply
% \begin{equation}
%     \frac{\rd v_{i,j+1/2}}{\rd t} = (\ldots) + \frac{1}{2} (T_{i,j} + T_{i,j+1}).
% \end{equation}
% This leads to an additional source term in the kinetic energy equation:
% \begin{equation}
%     \frac{\rd E_{k,h}}{\rd t} = - \mu \Phi + v_{h}^T ( A T_{h}),
% \end{equation}
% with $ A$ an averaging matrix.

Note that $\Phi_{h}$ is derived based on local energy consideration which upon summation equals the global dissipation, just like equation \eqref{eqn:epsU_phi}:
\begin{equation}\label{eqn:Phi_global}
    1^{T} \Omega_{p} \Phi_{h} = \epsilon_{U,h}.
\end{equation}

\subsection{Internal energy equation}
% Using in the discretization of the internal energy equation 
Having proposed a consistent expression for $\Phi_{h}$, the spatial discretization of the internal energy equation \eqref{eqn:T_Boussinesq_nondim3} reads:
% The discretised temperature equation reads:
\begin{equation}\label{eqn:temperature_equation_semidiscrete}
    \Omega_{p} \frac{\rd T_{h}}{\rd t} = -  C_{T}(V_{h},T_{h}) + \alpha_{3} \Omega_{p} \Phi_{h}(V_{h}) + \alpha_{4} ( D_{T} T_{h} + \hat{y}_{T}),
\end{equation}
% \begin{multline}\label{eqn:temperature_equation_discretized}
%     \Omega_{i,j} \frac{\rd T_{i,j}}{\rd t} = - \Delta y \left( u_{i+1/2,j} \frac{1}{2} (T_{i+1,j} + T_{i,j}) - u_{i-1/2,j} \frac{1}{2} (T_{i,j} + T_{i-1,j}) \right) \\ - \Delta x \left( v_{i,j+1/2} \frac{1}{2} (T_{i,j+1} + T_{i,j}) - v_{i,j-1/2} \frac{1}{2} (T_{i,j} + T_{i,j-1}) \right) + \\  \alpha_{4} \Delta y \left(  \frac{T_{i+1,j} - T_{i,j}}{\Delta x} -   \frac{T_{i,j} - T_{i-1,j}}{\Delta x} \right)  + \alpha_{4} \Delta x \left(  \frac{T_{i,j+1} - T_{i,j}}{\Delta y} -   \frac{T_{i,j} - T_{i,j-1}}{\Delta y} \right) 
%     + \alpha_{3} \Phi_{i,j} \Omega_{i,j},
% \end{multline}
where
\begin{equation}\label{eqn:internal_energy_convection}
    \begin{split}
        [ C_{T}(V_{h},T_{h})]_{i,j} = &\Delta y \left( u_{i+1/2,j} \frac{1}{2} (T_{i+1,j} + T_{i,j}) - u_{i-1/2,j} \frac{1}{2} (T_{i,j} + T_{i-1,j}) \right) + \\
        &\Delta x \left( v_{i,j+1/2} \frac{1}{2} (T_{i,j+1} + T_{i,j}) - v_{i,j-1/2} \frac{1}{2} (T_{i,j} + T_{i,j-1}) \right)
    \end{split}
\end{equation}
is the convection operator. The convection operator has a discrete skew-symmetry property which will be used in the derivation of the thermal dissipation balance in the next subsection. $D_{T}$ the standard second-order difference stencil with boundary conditions encoded in $\hat{y}_{T}$.
% equation \eqref{eqn:local_dissipation_function} 
% or in short
% \begin{equation}
%     \Omega \frac{\rd T_{h}}{\rd t} = - C^{T}_h (V_{h},T_{h}) + D^{T}_{h} T_{h} + \alpha_{3} \Omega \Phi + \text{boundary conditions}.
% \end{equation}

The total internal energy is given by $E_{i,h} = 1^{T} \Omega_{p} T_{h}$ (simply summing over all finite volumes). Due to the no-slip boundary conditions on the velocity field, the convective operator satisfies $1^{T} C_{T}(V_h,T_h) = 0$. The summation over the diffusive operator can be written in terms of the Nusselt numbers (detailed in the next section).
The total internal energy equation thus reads
\begin{equation}
\begin{split}
\frac{\rd E_{i,h}}{\rd t} &= \alpha_{3} 1^{T} \Omega_{p} \Phi_{h} + \alpha_{4} 1^{T}(D_{T} T_{h} + \hat{y}_{T}),\\
&=\alpha_{3} 1^{T} \Omega_{p} \Phi_{h} + \alpha_{4} (\Nu_{H} - \Nu_{C}),
\end{split}
\end{equation}
where in the second line the Nusselt numbers are instantaneous Nusselt numbers.
Upon adding the total kinetic energy equation \eqref{eqn:kineticenergy_semidiscrete}, and using property \eqref{eqn:Phi_global}, the global energy balance results:
\begin{equation}
\begin{split}
    \frac{\rd E_{h}}{\rd t} =  \frac{\rd E_{k,h}}{\rd t} + \gamma \frac{\rd E_{i,h}}{\rd t} &= \alpha_{2} V_{h}^T ( A T_{h} + y_{T}) +  \gamma \alpha_{4} 1^{T}(D_{T} T_{h} + \hat{y}_{T}),\label{eqn:totalenergy_balance_semidiscrete} \\
     &= \alpha_{2} V_{h}^T ( A T_{h} + y_{T}) + \gamma \alpha_{4} (\Nu_{H} - \Nu_{C}),
\end{split}
\end{equation}
which is the semi-discrete counterpart of equation \eqref{eqn:totalenergy_balance_nondim}. In other words, we have proposed a discrete viscous dissipation function that leads to a correct expression for the total energy equation, namely such that the viscous dissipation from the kinetic and internal energy equations exactly balances. Note that in the case of homogeneous Neumann boundary conditions for the temperature on all boundaries, the last term disappears. % since  \eqref{eqn:summation_diffusion} evaluates to zero.

\subsection{Discrete global balances and Nusselt number relations}\label{sec:Nusselt_discretization}
We now derive discrete versions of the Nusselt relations that incorporate the viscous dissipation function, i.e.\ relations \eqref{eqn:Nusselt_epsU} and \eqref{eqn:Nusselt_thermal_diss}. Our symmetry-preserving spatial discretization is such that \textit{exact} discrete relations can be derived. It is important to realize that the discrete approximation for the Nusselt number cannot be chosen independently (when the goal is to have exact discrete global balances) but is implicitly defined once the discretization of the diffusive operator is chosen. Consider the discretized global internal energy equation for steady conditions, 
\begin{equation}
    \alpha_{3} 1^{T} \Omega_{p} \Phi_{h}(V_{h}) + \alpha_{4} 1^{T} ( D_{T} T_{h} + \hat{y}_{T})= 0.
\end{equation}
The second term can be simplified as
\begin{equation}\label{eqn:summation_diffusion}
    1^{T} ( D_{T} T_{h} + \hat{y}_{T}) = - \sum_{i=1}^{N_{x}} \frac{T_{i,1}-T_{H}}{\frac{1}{2}\Delta y} \Delta x +  \sum_{i=1}^{N_{x}} \frac{T_{C} - T_{i,N_{y}}}{\frac{1}{2}\Delta y} \Delta x = \Nu_{H} - \Nu_{C},
\end{equation}
where the Nusselt numbers on the lower (hot) and upper (cold) plate are defined as
\begin{align}
    \Nu_{H} &:= - \sum_{i=1}^{N_{x}} \frac{T_{i,1} - T_{H}}{\frac{1}{2}\Delta y} \Delta x, \label{eqn:Nusselt_H} \\
    \Nu_{C} &:= - \sum_{i=1}^{N_{x}} \frac{T_{C} - T_{i,N_{y}}}{\frac{1}{2}\Delta y} \Delta x. \label{eqn:Nusselt_C}
\end{align}
This leads to the discrete version of \eqref{eqn:Nusselt_epsU}:
\begin{equation}\label{eqn:Nusselt_diff}
     \alpha_{4} (\Nu_{C} - \Nu_{H}) = \alpha_{3} 1^{T} \Omega_{p} \Phi_{h}(V_{h}).
\end{equation}

The discrete version of \eqref{eqn:Nusselt_thermal_diss} follows by considering the inner product of equation \eqref{eqn:temperature_equation_semidiscrete} with $T_{h}^T$ instead of $1^{T}$. An important property of the convective discretization \eqref{eqn:internal_energy_convection} is that
\begin{equation}
    T_{h}^T C_{T} (V_{h},T_{h}) = 0, \qquad \forall \, T_{h}, \quad \mathrm{if} \quad M V_{h}=0.
\end{equation}
This property is most easily derived by recognizing that $C_{h} (V_{h},T_{h})$ can be written in terms of a matrix-vector product $\tilde{C}_{T} (V_{h})T_{h}$, where $\tilde{C}_{T} (V_{h})$ is skew-symmetric if $M V_{h}=0$. In addition, the inner product of $T_{h}$ with the diffusive terms can be written as 
\begin{equation}\label{eqn:discrete_dissipation_temperature}
    T_{h}^T (D_{T} T_{h} + \hat{y}_{T}) = \sum_{i=1}^{N_{x}} \left(- T_{H}\frac{T_{i,1} - T_{H}}{\frac{1}{2}\Delta y} + T_{C} \frac{T_{C} - T_{i,N_{y}}}{\frac{1}{2}\Delta y}  \right)\Delta x -\epsilon_{T,h},
\end{equation}
where
\begin{equation}
\begin{split}
   \epsilon_{T,h} :=&  \sum_{i=1}^{N_{x}} \left( \frac{1}{2} \left(\frac{T_{i,1} - T_{H}}{\frac{1}{2}\Delta y} \right)^2 + \sum_{j=2}^{N_{y}} \left( \frac{T_{i,j}-T_{i,j-1}}{\Delta y} \right)^2 + \frac{1}{2} \left(\frac{T_{C} - T_{i,N_{y}} }{\frac{1}{2}\Delta y} \right)^2 \right) \Delta x \Delta y  + \sum_{j=1}^{N_{y}} \sum_{i=2}^{N_{x}} \left( \frac{T_{i,j}-T_{i-1,j}}{\Delta x} \right)^2 \Delta x \Delta y
\end{split}
\end{equation}
is the discrete analogue of \eqref{eqn:thermal_dissipation} and equation \eqref{eqn:discrete_dissipation_temperature} is the discrete version of $\int T \frac{\rd^2 T}{\rd y^2} =  [T \frac{\rd T}{\rd y}] -\int (\frac{\rd T}{\rd y})^2$. With the boundary condition $T_{H}=1$, $T_{C}=0$, we get the balance
\begin{equation}\label{eqn:Nusselt_thermal_diss_discrete}
     \alpha_{4} \Nu_{H} = \alpha_{4} \epsilon_{T,h} - \alpha_{3} T_{h}^{T} \Omega_{p} \Phi_{h}(V_{h}),
\end{equation}
which is the discrete version of equation \eqref{eqn:Nusselt_thermal_diss}.

\section{Energy-consistent temporal discretization}\label{sec:temporal_discretization}
The system of equations \eqref{eqn:mass_semidiscrete}, \eqref{eqn:mom_semidiscrete} and \eqref{eqn:temperature_equation_semidiscrete} needs to be integrated in time with a suitable method in order to preserve a time-discrete version of the global energy balance \eqref{eqn:totalenergy_balance_semidiscrete}. A common choice is to use an explicit method (e.g.\ Adams-Bashforth) for the nonlinear convective terms and an implicit method (e.g.\ Crank-Nicolson) for the (stiff) linear diffusion terms \cite{gayen2013,hepworth2014,sugiyama2009}, or an explicit method for both convection and diffusion \cite{verstappen2003, trias2011}. In such an approach, the temperature equation is typically solved first (given velocity fields at previous time instances), and then the mass and momentum equations are solved with a pressure-correction approach. However, these methods do not preserve the global energy balance as they violate the energy-conserving nature of the nonlinear terms when marching in time \cite{sanderse2013}. 

Instead, we show here that the implicit midpoint method can be employed to achieve energy-consistent time integration. The fully discrete system reads:
\begin{align}
     M V_{h}^{n+1/2} &= 0, \label{eqn:mass_fullydiscrete} \\
    \Omega_{V} \frac{V^{n+1}_{h} - V^{n}_{h}}{\Delta t} &= - C_{V} (V_{h}^{n+1/2}) -  G p_{h}^{n+1/2} +  \alpha_{1} D_{V} V_{h}^{n+1/2} +  \alpha_{2} (A T_{h}^{n+1/2} + y_T), \label{eqn:mom_fullydiscrete} \\
    \Omega_{p} \frac{T^{n+1}_{h} - T^{n}_{h}}{\Delta t} &= -C_{T} (V_{h}^{n+1/2},T_{h}^{n+1/2}) +  \alpha_{3} \Omega_{p} \Phi (V_{h}^{n+1/2}) + \alpha_{4} (D_{T} T_{h}^{n+1/2} + \hat{y}_{T}). \label{eqn:internalenergy_fullydiscrete}
\end{align}
Here $V_{h}^{n+1/2} = \frac{1}{2} (V_{h}^{n} + V_{h}^{n+1})$ and $T_{h}^{n+1/2} = \frac{1}{2} (T_{h}^{n} + T_{h}^{n+1})$. Upon multiplying \eqref{eqn:mom_fullydiscrete} by $(V_{h}^{n+1/2})^{T}$ and \eqref{eqn:internalenergy_fullydiscrete} by $1^{T}$, and adding the two resulting equations, we get the discrete energy balance, 
\begin{equation}
    \frac{E_{h}^{n+1} - E_{h}^{n}}{\Delta t} = \frac{E_{k,h}^{n+1} - E_{k,h}^{n}}{\Delta t} + \gamma \frac{E_{i,h}^{n+1} - E_{i,h}^{n}}{\Delta t} = \alpha_{2} (V_{h}^{n+1/2})^T ( A T_{h}^{n+1/2} + y_{T}) +  \gamma \alpha_{4} 1^{T}(D_{T} T_{h}^{n+1/2} + \hat{y}_{T}),
\end{equation}
which is the fully-discrete counterpart of equations \eqref{eqn:totalenergy_balance_nondim} and \eqref{eqn:totalenergy_balance_semidiscrete}. The derivations hinges again on skew-symmetry of the convection operator $C_{V}$, the compatibility between $M$ and $G$ ($G=-M^T$), and the consistency requirement on the viscous dissipation function, equation \eqref{eqn:Phi_global}.

The system of equations \eqref{eqn:mass_fullydiscrete} - \eqref{eqn:internalenergy_fullydiscrete} leads to a large system of nonlinear equations which has a saddle point structure due to the divergence-free constraint. We solve the system in a segregated fashion and iterate at each time step with a standard pressure-correction method until the residual of the entire system is below a prescribed tolerance. We will compare this energy-conserving time integration approach to an explicit one-leg method \cite{verstappen2003,trias2011} in section \ref{sec:results_timedependent}.

\section{Steady state results (Rayleigh-Bénard)}\label{sec:results_steady}
The concept of energy consistency is best demonstrated through time-dependent simulations. However, we start with steady-state results in order to validate the spatial discretization method and to get intuition for the effect of the Gebhart number on the Nusselt number. For the results reported here we employ a direct solver that solves the entire coupled non-linear system of equations that arises from spatial discretization. As initial guess we take the following divergence-free velocity field:
\begin{align}
    u(x,y) &= -64 x^2 (x-1)^2 y (y-1)(2y-1), \\
    v(x,y) &= 64 x (x-1)(2x-1) y^2 (y-1)^2, 
\end{align}
which is inspired by the regularized driven cavity problem \cite{shih1989}. For the temperature we take a random field (between 0 and 1). The idea behind this choice of initial condition is to avoid the non-linear solver to be stuck in the trivial solution ($\vt{u}=0$). Note that in all simulations in this article, we will set $\Pr=0.71$ (air), and use non-dimensionalization choice I. Choices II and III give equivalent results apart from scaling factors. 
% and we resort to a standard Adams-Bashforth Crank-Nicolson (AB-CN) scheme to quickly march towards a steady state.

\subsection{Grid convergence study for no-dissipation case (Ge = 0)}

\begin{comment}
  We integrate the RBC problem for $\Ra = 10^{4}$ and $\Pr=0.71$ with AB-CN and $\Delta t = 1/10$ until $t_{\text{end}}=200$. We take $T_{H} = 1$ and $T_{C} = 0$ and the initial condition for $\vt{u}$ is homogeneous, for $T$ we take $1-y$. Note that only after $t \approx 100$ the RB instability appears, and the transition to a new steady state takes place. At $t=200$ the residual of the momentum and energy equation is at machine precision. We perform a grid convergence study and compare our results with literature in terms of the Nusselt number on the hot and cold plate, defined by equations \eqref{eqn:Nusselt_H} and \eqref{eqn:Nusselt_C}.
% \begin{equation}\label{eqn:Nu_H}
%     \Nu_{H} = - \int_{0}^{1} \left(\dd{T}{y}\right)_{y=0} \rd x \approx - \sum_{i=1}^{N_x}  \frac{ -\frac{1}{3} T_{i,2} + 3 T_{i,1} - \frac{8}{3} T_{H}}{\Delta y} \Delta x.
% \end{equation}
% Note however, that in order to get the heat flux that is consistent with the discretization of the (diffusive term in the) temperature equation, one should use a first order approximation.

For finer mesh or higher $\Ra$ we decrease the time step according to the convective stability requirement. For example, for $\Ra = 10^{5}$ we take $\Delta t = 2/100$.
\end{comment}
Figure \ref{fig:RBC_Temp_Ra1e5_Ge0_N128} shows the temperature field when viscous dissipation is not included ($\Ge = 0$). The resulting Nusselt numbers as a function of grid refinement are displayed in Table \ref{tab:Nu_grid_convergence} and indicate excellent agreement with literature \cite{cai2019}. We note that the Nusselt numbers as defined by \eqref{eqn:Nusselt_H} and \eqref{eqn:Nusselt_C} are first-order approximations. More accurate approximations can be constructed by including more interior points. We are not using such high-order accurate approximations as they would not satisfy the discrete global balance \eqref{eqn:Nusselt_diff}. Note also that we only report $\Nu_{H}$ since $\Nu_{C} = \Nu_{H}$ up to machine precision.

% \begin{figure}[h]
%     \centering
%     \includegraphics[width=0.4\textwidth]{Ra1e4_64x64.pdf}
%     \caption{Steady-state temperature and flow field for $\Ra = 10^{4}$ on a 64 $\times$ 64 grid.}
%     \label{fig:Ra1e4_64x64}
% \end{figure}
\begin{figure}[ht]
\begin{subfigure}[b]{0.32\textwidth}
    \centering
    \includegraphics[width=\textwidth]{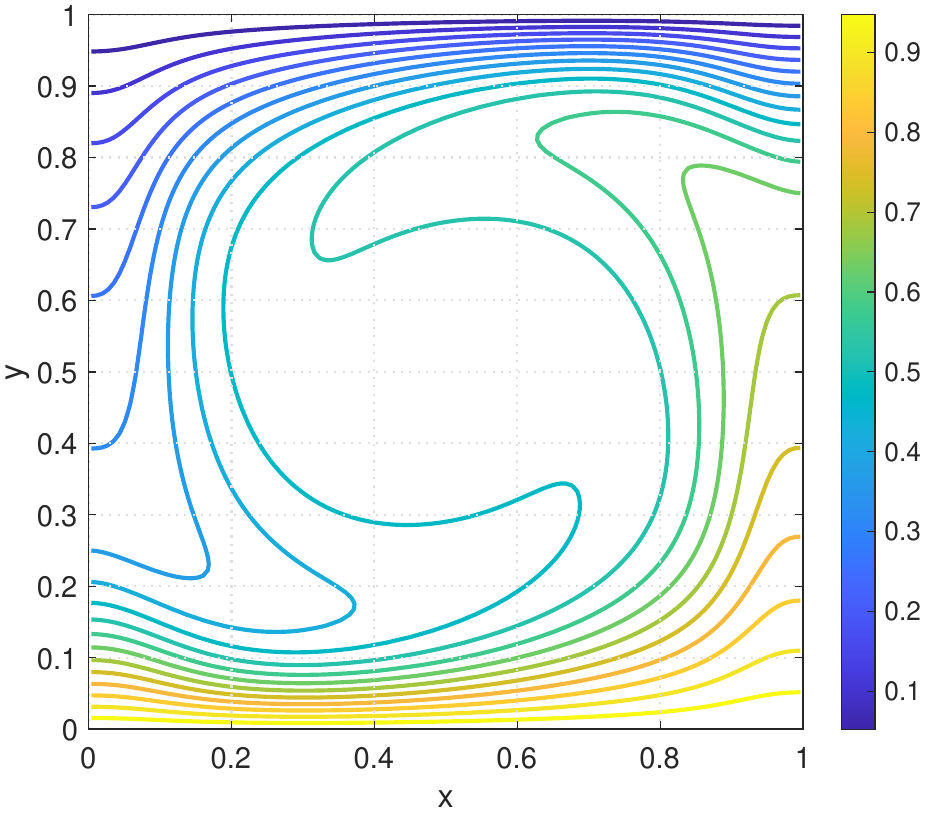}
    \caption{$\Ge = 0$.}
    \label{fig:RBC_Temp_Ra1e5_Ge0_N128}
\end{subfigure}
\hfill
\begin{subfigure}[b]{0.32 \textwidth}
    \includegraphics[width=\textwidth]{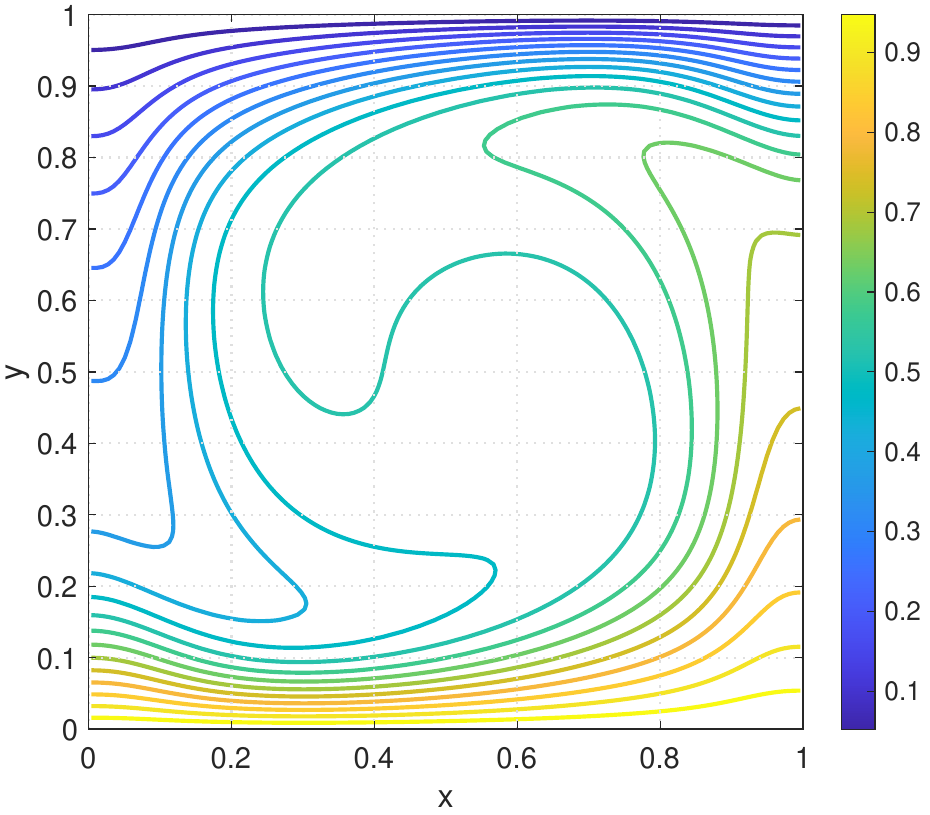}
    \caption{$\Ge = 0.1$.}
    \label{fig:RBC_Temp_Ra1e5_Ge1e-1_N128}
\end{subfigure}
\hfill
\begin{subfigure}[b]{0.32 \textwidth}
    \includegraphics[width=\textwidth]{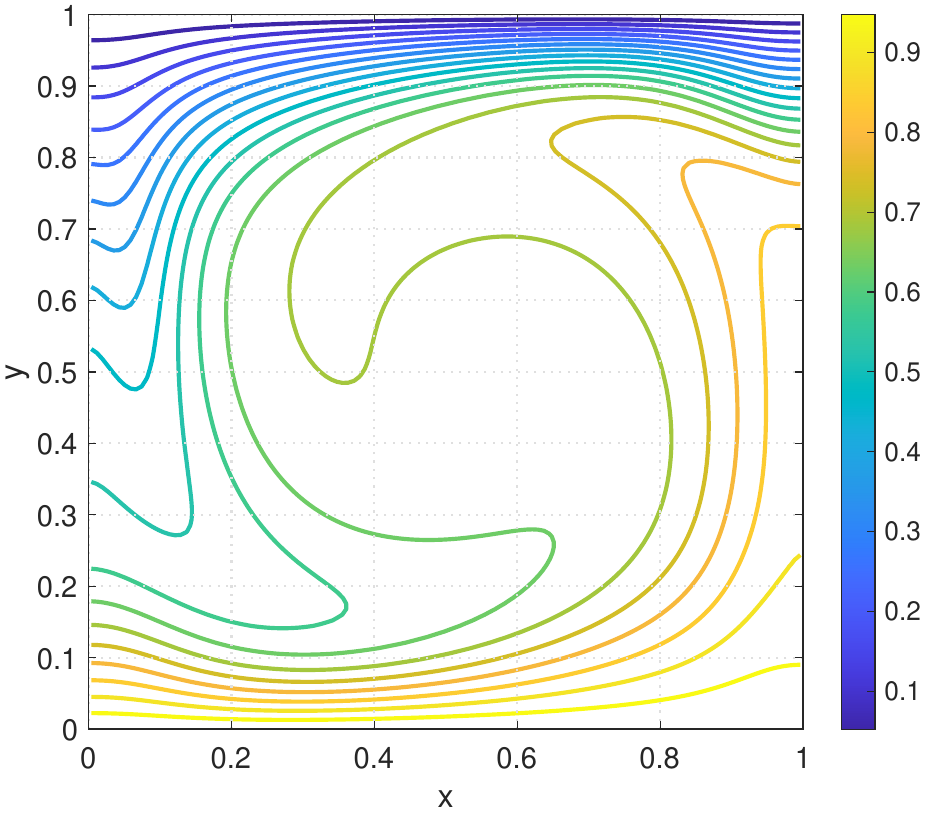}
    \caption{$\Ge = 1$.}
    \label{fig:RBC_Temp_Ra1e5_Ge1e0_N128}
\end{subfigure}
\caption{Steady-state temperature field for $\Ra = 10^{5}$ on a 128 $\times$ 128 grid, for different $\Ge$.
    \label{fig:RBC_Temp_Ra1e5_Ge_N128}}
\end{figure}

% \begin{figure}[h]
% \begin{subfigure}[b]{0.45 \textwidth}
%      \centering
%     \includegraphics[width=\textwidth]{Nu_time_Ra_64x64.pdf}
%     \caption{Without dissipation.}
%     \label{fig:Nu_time_Ra_64x64}    
% \end{subfigure}
% \hfill
% \begin{subfigure}[b]{0.45 \textwidth}
% \centering
%     \includegraphics[width=\textwidth]{Nu_time_Ra1e4_64x64_withdissipation.pdf}
%     \caption{With dissipation, $\Ra=10^{4}$, $\Ge =1$.}
%     \label{fig:Nu_time_Ra1e4_64x64_withdissipation}    
% \end{subfigure}
%     \caption{Convergence of Nusselt number \eqref{eqn:Nusselt_H} to steady state for different $\Ra$ on a 64 $\times$ 64 grid.}
% \end{figure}

% \begin{table}[h]
%     \centering
%     \begin{tabular}{l c c c  c c}
%     \toprule
%          & \multicolumn{3}{c}{without dissipation} &  \multicolumn{2}{c}{with dissipation}  \\
%         grid & $\Ra= 10^3$ & $\Ra= 10^4$ & $\Ra= 10^5$ & $\Ra= 10^{4}$, $\Ge=1$ & $\Ra= 10^{4}$, $\Ge=0.1$\\
%     \midrule
%        $32^2$  & 1.000 & 2.172 &   3.959 & 1.563 & 2.111 \\
%        $64^2$  & 1.000 &  2.161 & 3.920 & 1.567 & 2.102 \\
%        $128^2$ & 1.000 & 2.159 & 3.912 & 1.571 & 2.100 \\
%       $256^2$ & 1.000 & 2.158 & 3.911 & 1.573 & 2.100 \\
%        Cai et al.\ \cite{cai2019} ($256^2$) & 1.000 & 2.158 & 3.911 & - & -\\
%      \bottomrule
%     \end{tabular}
%     \caption{Convergence of Nusselt number \eqref{eqn:Nu_H} with grid refinement for different Rayleigh numbers and Gebhart numbers.}
%     \label{tab:Nu_grid_convergence}
% \end{table}

\begin{table}[ht]
    \centering
    \begin{tabular}{l c c c }
    \toprule
        grid & $\Ra= 10^3$ & $\Ra= 10^4$ & $\Ra= 10^5$ \\
    \midrule
       $32^2$  & 1.000 & 2.170 & 3.933 \\ 
       $64^2$  & 1.000 &  2.161 & 3.916 \\
       $128^2$ & 1.000 & 2.159 & 3.912 \\
      $256^2$ & 1.000 & 2.158 & 3.911 \\
       Cai et al.\ \cite{cai2019} ($256^2$) & 1.000 & 2.158 & 3.911\\
     \bottomrule
    \end{tabular}
    \caption{Convergence of Nusselt number \eqref{eqn:Nusselt_H} with grid refinement for different Rayleigh numbers and $\Ge=0$.}
    \label{tab:Nu_grid_convergence}
\end{table}

\subsection{Grid convergence study for viscous dissipation case (Ge> 0)}
When including viscous dissipation ($\Ge > 0$) in the internal energy equation, the flow field changes qualitatively and loses its symmetric nature, as can be observed in figures \ref{fig:RBC_Temp_Ra1e5_Ge1e-1_N128}-\ref{fig:RBC_Temp_Ra1e5_Ge1e0_N128}. The Nusselt numbers at the hot and cold plate start to deviate from each other, their difference being equal to the dissipation function, according to equation \eqref{eqn:Nusselt_diff} (or \eqref{eqn:Nusselt_epsU}). This is reported in table \ref{tab:Nu_Ge_grid_convergence} and figure \ref{fig:Ra_Nu_bifurcation_annotated}. The critical Rayleigh number that we find from the bifurcation diagram is $\Ra_{c}\approx 2585$, which is in excellent agreement with the value of 2585.02 reported in literature \cite{gelfgat1999,venturi2010}. It is independent of the value of the Prandtl number, as shown in \cite{gelfgat1999}, and also independent of the value of the Gebhart number. This latter fact follows by extending the linear stability analysis in \cite{gelfgat1999} and realizing that the term $\nabla \vt{u}:\nabla \vt{u}$ with $\vt{u} = \vt{u}_{0} + \varepsilon \vt{u}'$ and background state $\vt{u}_0=0$ leads to the term $\varepsilon^2 \nabla \vt{u}' :\nabla \vt{u}'$, which disappears when gathering terms of $\mathcal{O}(\varepsilon)$. The results in figure \ref{fig:Ra_Nu_bifurcation_annotated} show indeed that the bifurcation point is the same for different values of $\Ge$.

Figure \ref{fig:Ra_Nu_bifurcation_epsT_annotated} shows a different interpretation of the Nusselt number, indicating the relation with the thermal dissipation and viscous dissipation according to equation \eqref{eqn:Nusselt_thermal_diss_discrete} (or \eqref{eqn:Nusselt_thermal_diss}). The results confirm that the thermal dissipation lies in between the Nusselt number of the hot and cold plate.

% We get $\Nu = 2.172$ (32x32), $\Nu = 2.161$ (64x64), $\Nu = 2.159$ (128x128).
% For $\Ra = 10^{5}$, $\Nu = 3.959$ (32x32), $\Nu = 3.920$ (64x64), $\Nu = 3.912$ (128x128).
% For $\Ra = 10^{3}$, $\Nu = 1.000$ for all resolutions.

\begin{table}[ht]
    \begin{subtable}[h]{0.45\textwidth}
    \centering
        \begin{tabular}{l c c c c}
        \toprule
            grid &  \multicolumn{2}{c}{$\Ge=0.1$} & \multicolumn{2}{c}{$\Ge=1$} \\
            & $\Nu_{H}$ & $\Nu_{C}$ &  $\Nu_{H}$ & $\Nu_{C}$  \\        
        \midrule
           $32^2$ & 2.111 & 2.228 & 1.582 & 2.729 \\
           $64^2$  &  2.103 & 2.219 & 1.578 & 2.716 \\ 
           $128^2$ & 2.101 & 2.217 & 1.576 & 2.713\\
          $256^2$  & 2.100 & 2.216 & 1.576 & 2.712 \\
         \bottomrule
        \end{tabular}
        \label{tab:Nu_Ge_Ra1e4_grid_convergence}
        \caption{$\Ra = 10^{4}$.}
    \end{subtable}
    \hfill
    \begin{subtable}[h]{0.45\textwidth}
        \begin{tabular}{l c c c c}
        \toprule
            grid &  \multicolumn{2}{c}{$\Ge=0.1$} & \multicolumn{2}{c}{$\Ge=1$} \\
            & $\Nu_{H}$ & $\Nu_{C}$ &  $\Nu_{H}$ & $\Nu_{C}$  \\        
        \midrule
           $32^2$ & 3.786 & 4.080 & 2.448 & 5.319 \\
           $64^2$  &  3.770 & 4.062 & 2.441 & 5.299 \\
           $128^2$ & 3.766 & 4.057 & 2.439 & 5.293 \\
          $256^2$  & 3.765 & 4.056 & 2.439 & 5.292 \\
         \bottomrule
        \end{tabular}
        \label{tab:Nu_Ge_Ra1e5_grid_convergence}
        \caption{$\Ra = 10^{5}$.}
    \end{subtable}
    \caption{Convergence of Nusselt numbers \eqref{eqn:Nusselt_H} and \eqref{eqn:Nusselt_C} with grid refinement for different Rayleigh and different Gebhart numbers.
    \label{tab:Nu_Ge_grid_convergence}}
    % \label{tab:Nu_Ge_Ra1e4_grid_convergence}
\end{table}

\begin{figure}[hbtp]
\centering 
\begin{subfigure}[t]{0.42 \textwidth}
     \centering
\includegraphics[width=\textwidth]{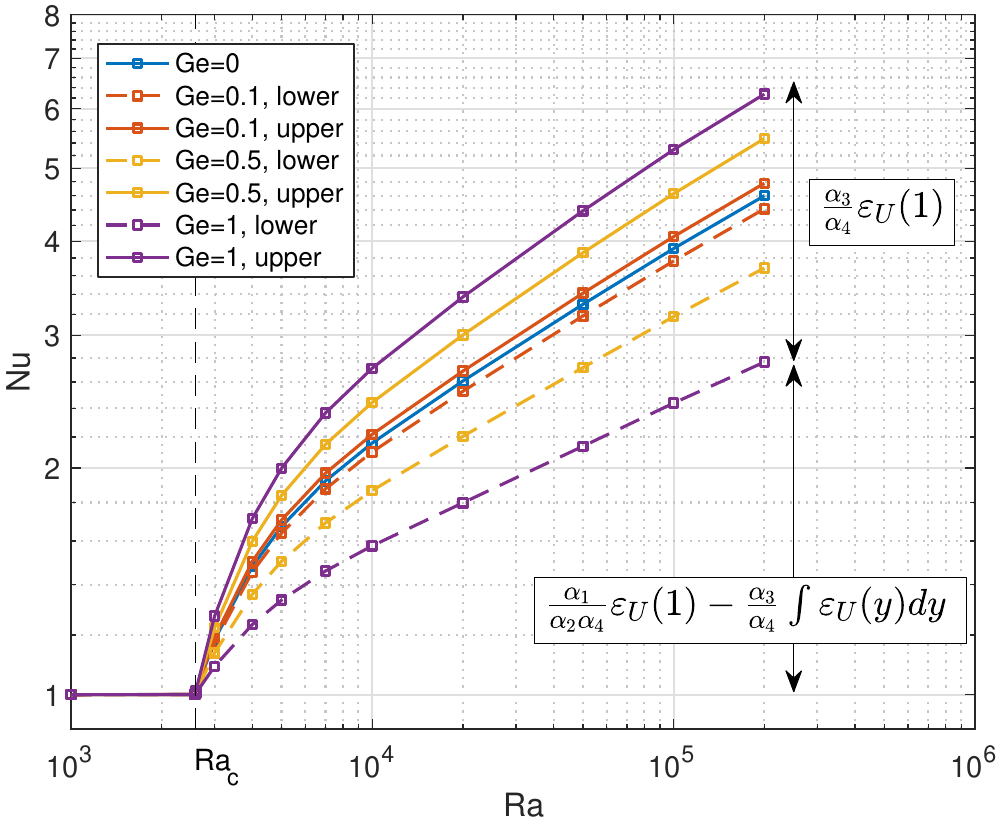} 
\caption{Viscous dissipation does not change the critical Rayleigh number, but leads to a difference between the Nusselt number of the upper and lower plate.}
\label{fig:Ra_Nu_bifurcation_annotated}
\end{subfigure}  
\hfill
\begin{subfigure}[t]{0.49 \textwidth}
     \centering
    \includegraphics[width=\textwidth]{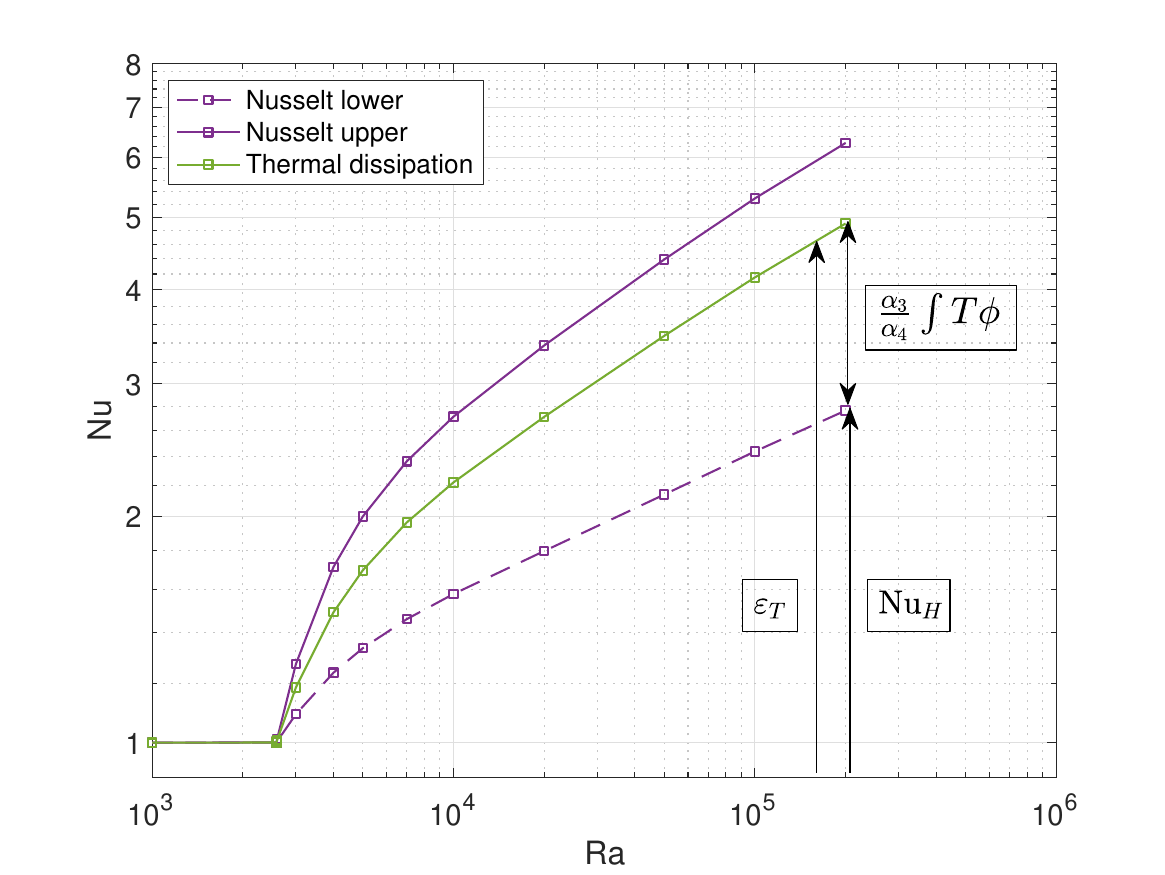}
    \caption{Relation between thermal dissipation and viscous dissipation for $\Ge=1$.}
    \label{fig:Ra_Nu_bifurcation_epsT_annotated}    
\end{subfigure}  
\caption{Bifurcation diagram for Rayleigh-Bénard problem including viscous dissipation.}
\end{figure}

\FloatBarrier

\section{Time-dependent, energy-conserving simulation (Rayleigh-Taylor)}\label{sec:results_timedependent}
% \subsection{Energy conservation}
The previous section confirmed the (discrete) steady-state Nusselt number balances. In this section we consider the core idea of this article: achieving exact energy conservation in a time-dependent simulation. Exact energy conservation requires that all contributions from boundary terms disappear, which we achieve by prescribing no-slip conditions $\vt{u}=0$ and adiabatic conditions $\dd{T}{n}=0$ on all boundaries (the pressure does not require boundary conditions). The energy balance then represents a pure exchange of kinetic, internal and potential energy according to 
\begin{equation}\label{eqn:energy_exchange_results}
    \frac{E_{h}^{n+1} - E_{h}^{n}}{\Delta t} = \frac{E_{k,h}^{n+1} - E_{k,h}^{n}}{\Delta t} + \gamma \frac{E_{i,h}^{n+1} - E_{i,h}^{n}}{\Delta t} = \alpha_{2} (V_{h}^{n+1/2})^T ( A T_{h}^{n+1/2} + y_{T}).
\end{equation}
However, with adiabatic boundary conditions we cannot simulate the classic Rayleigh-Bénard problem. Instead, we turn to the well-known Rayleigh-Taylor problem, featuring a cold (heavy) fluid on top of a warm (light) fluid. A sketch of the set-up is shown in figure \ref{fig:problem_setup_RayleighTaylor}. The energy-conserving implicit midpoint (`IM') method detailed in section \ref{sec:temporal_discretization} will be compared to the explicit one-leg (`OL') method commonly used in DNS studies \cite{verstappen2003,trias2011} (where we take $\kappa=\frac{1}{2}$ and a fixed time step).

\begingroup
\begin{figure}[hbtp]
\fontfamily{lmss} % Latin Modern Sans Serif
%\fontfamily{phv} % helvetica
\fontsize{10pt}{10pt}\selectfont
\centering 
\def\svgwidth{0.5 \textwidth}
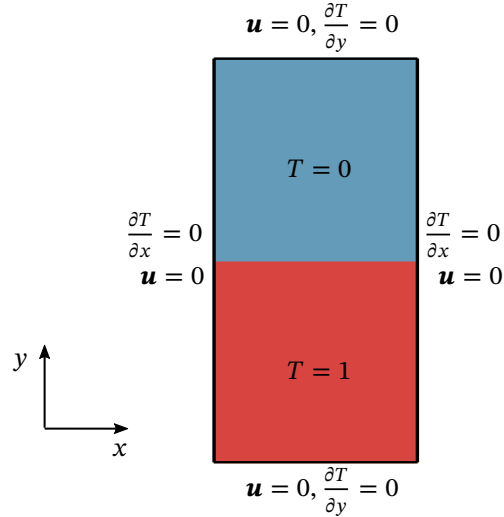 
\caption{Problem set-up with initial condition for Rayleigh Taylor problem.}
\label{fig:problem_setup_RayleighTaylor}
\end{figure}
\endgroup

The domain size is $1\times 2$, the grid is $64 \times 128$, the time step $\Delta t=5 \cdot 10^{-3}$ and the end time $T=100$. We consider the case $\Ra = 10^6$ and $\Ge = \{ 0.1, 1\}$ . The instability naturally arises due to growth of round-off errors (no perturbation is added in the initial condition). After the initial instability has developed, an asymmetry in the solution appears, triggering a sequence of well-known `mushroom' type plumes: hot plumes rising upward and cold plumes sinking downward (figure \ref{fig:RayleighTaylor_T}). Noteworthy are the differences in the development of the instability due to different time integration methods: IM predicts an earlier onset (around $t=23$) than OL (around $t=33$), and the evolution stays symmetric for a much longer period of time in case of IM. For both methods, the time of onset of instability is insensitive to the value of $\Ge$, just like the bifurcation point in the steady state Rayleigh-Bénard simulation was insensitive to the value of $\Ge$. The differences between the methods might be attributed to the absence of artificial dissipation in the IM scheme compared to OL, as well as its more symmetric nature. However, one should note that the problem is chaotic, and similar differences in the solution can also be obtained by adding minute perturbations to the initial condition. 
%The question `which time-integration scheme is the most accurate' can therefore not be answered easily with single simulation results, but would require an ensemble of simulations.
% This is confirmed in figure \ref{fig:results_RayleighTaylor}.

\begin{figure}[hbtp]
\centering 
\begin{subfigure}[b]{0.235 \textwidth}
     \centering
    \includegraphics[width=\textwidth]{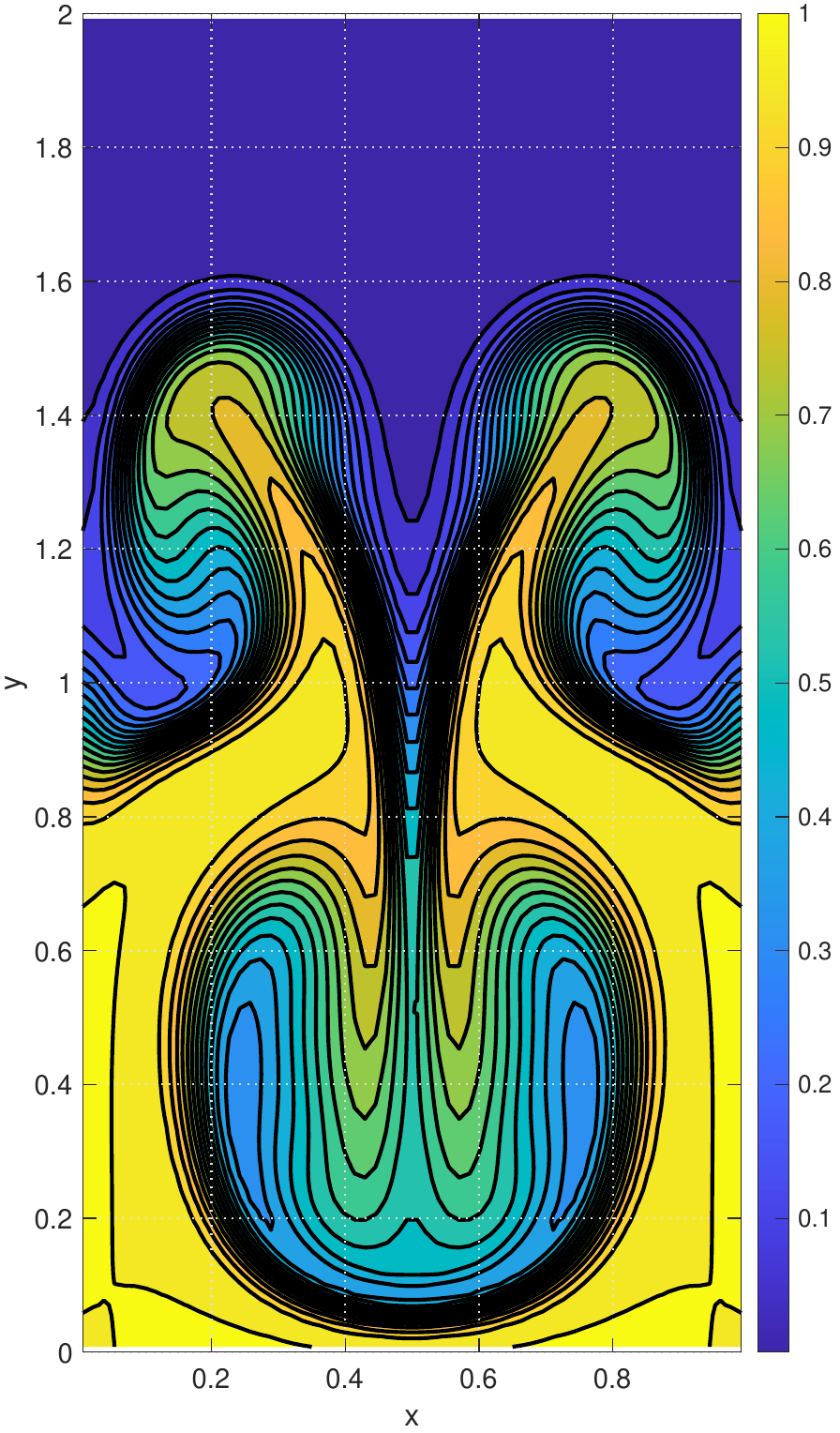}
    \caption{$t=30$, $\Ge=0.1$, IM}
    \label{fig:RayleighTaylor_T_method_23_Ge_0.1_t_30}    
\end{subfigure}
     \hfill
\begin{subfigure}[b]{0.24 \textwidth}
     \centering
    \includegraphics[width=\textwidth]{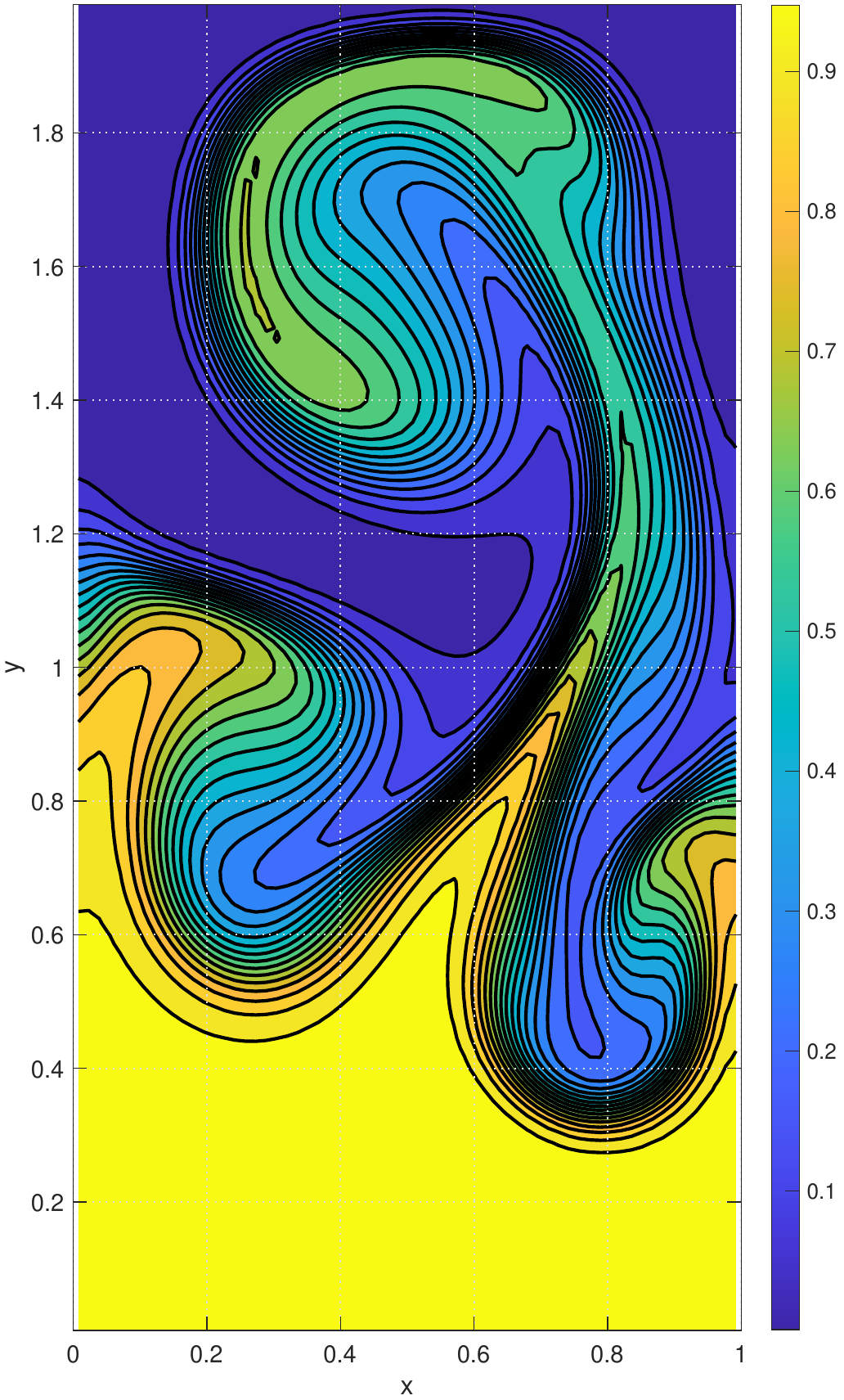}
    \caption{$t=40$, $\Ge=0.1$, OL}
    \label{fig:RayleighTaylor_T_method_5_Ge_0.1_t_40}    
\end{subfigure}    
\hfill
\begin{subfigure}[b]{0.245 \textwidth}
     \centering
    \includegraphics[width=\textwidth]{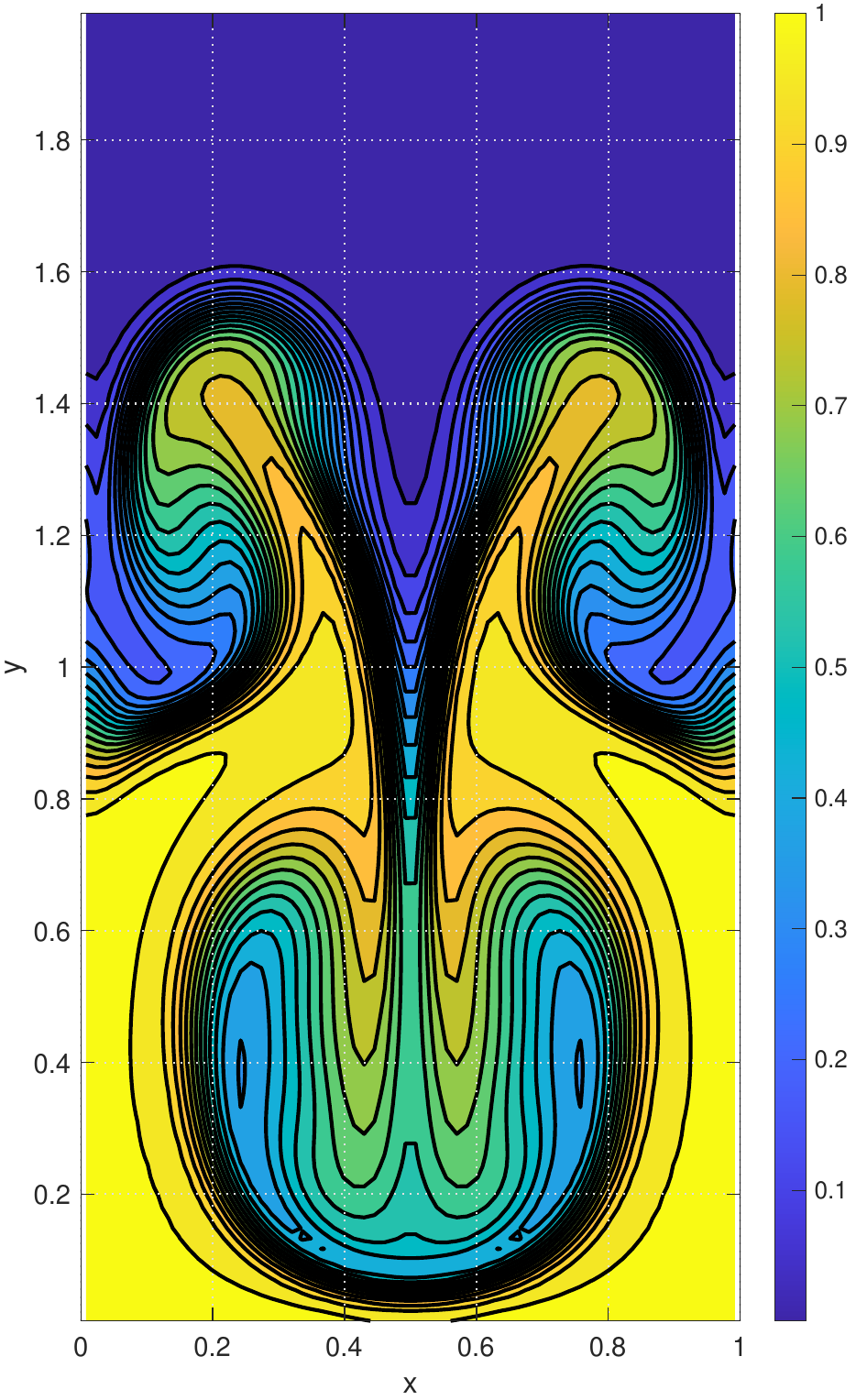}
    \caption{$t=30$, $\Ge=1$, IM}
    \label{fig:RayleighTaylor_T_method_23_Ge_1_t_30}    
\end{subfigure}
     \hfill
\begin{subfigure}[b]{0.25 \textwidth}
     \centering
    \includegraphics[width=\textwidth]{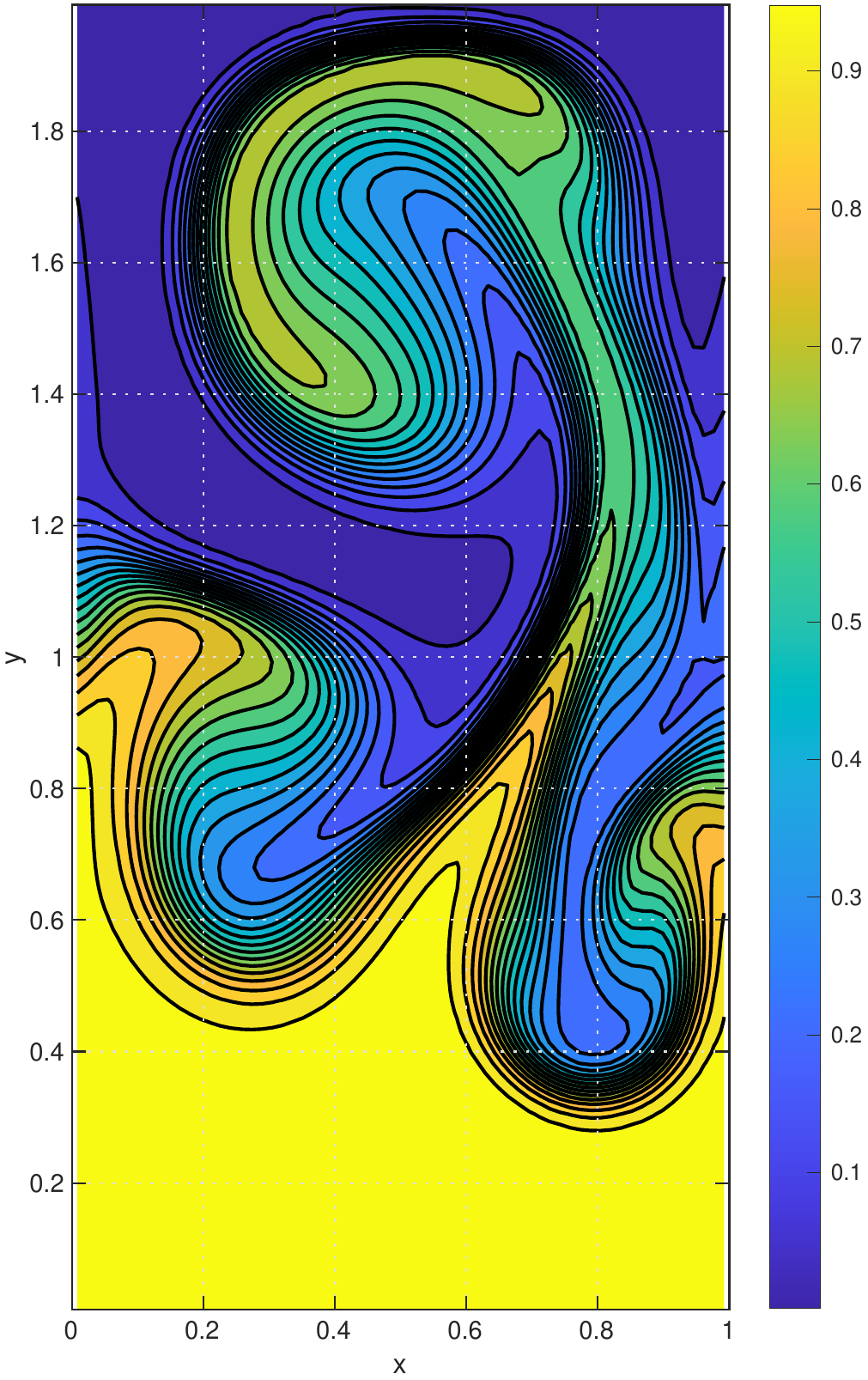}
    \caption{$t=40$, $\Ge=1$, OL}
    \label{fig:RayleighTaylor_T_method_5_Ge_1_t_40}    
\end{subfigure}    
    \caption{Rayleigh-Taylor temperature fields.}
\label{fig:RayleighTaylor_T}
\end{figure}

Since there is no driving force and all boundary conditions are homogeneous, viscosity damps the velocity field back to a homogeneous steady state, while at the same time increasing the temperature through dissipation. This increase in temperature is clear from figure \ref{fig:RayleighTaylor_T_avg_comparison}, where the average temperature is displayed. Compared to the initial temperature difference $\Delta T=1$, the relative temperature increase is about $2 \%$ for $\Ge=0.1$ and more than $20 \%$ for $\Ge = 1$. Note that many existing natural convection models, which ignore the viscous dissipation term, would not predict any temperature increase. With our proposed energy-consistent viscous dissipation function, the temperature increase exactly matches the kinetic energy loss through viscous dissipation. This is confirmed in figure \ref{fig:RayleighTaylor_dEdt_diff_Ra1e6_comparison}, which shows the energy error %in equation \eqref{eqn:energy_exchange_results}.
\begin{equation}
     \varepsilon_{E} := \left| \frac{E_{k,h}^{n+1} - E_{k,h}^{n}}{\Delta t} + \gamma \frac{E_{i,h}^{n+1} - E_{i,h}^{n}}{\Delta t} - \alpha_{2} (V_{h}^{n+1/2})^T ( A T_{h}^{n+1/2} + y_{T}) \right|.
\end{equation}
For IM the error remains at the tolerance with which we solve the system of nonlinear equations ($10^{-12}$). For OL, the error is initially at a similar level but increases to $\mathcal{O}(10^{-6})$ when the instability develops (for $t>30$). However, one could argue that this advantage is offset by the fact that IM is roughly 4-5$\times$ as expensive because it requires roughly 4-5 iterations (Poisson solves) per time step, making OL much faster to run. Consequently, OL will be employed for the 3D simulations in the next section. Note that this balance of accuracy versus computational costs depends on the details of the flow problem and might differ in other test cases.

\begin{figure}[hbtp]
\centering 
\begin{subfigure}[b]{0.45 \textwidth}
     \centering
    \includegraphics[width=\textwidth]{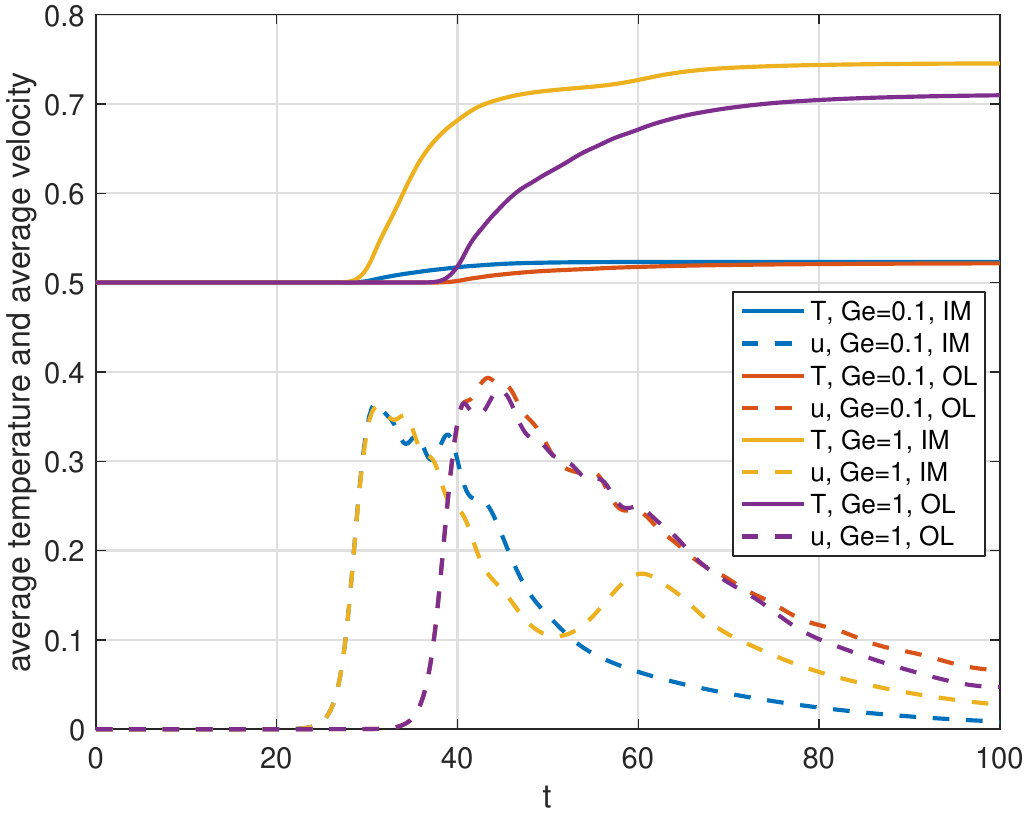}
    \caption{Average temperature and velocity.}
    \label{fig:RayleighTaylor_T_avg_comparison}    
\end{subfigure}
    \hfill
\begin{subfigure}[b]{0.5 \textwidth}
     \centering
    \includegraphics[width=\textwidth]{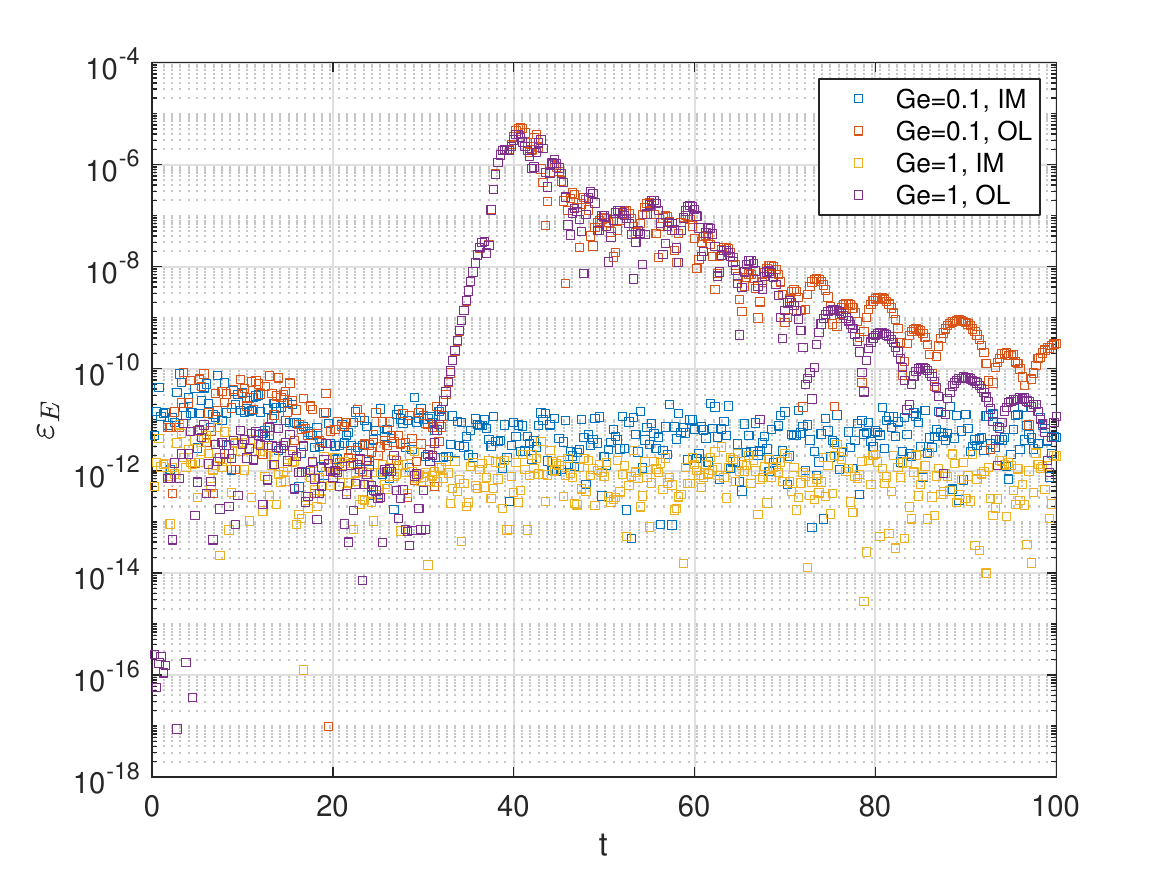}
    \caption{Energy error.}
    \label{fig:RayleighTaylor_dEdt_diff_Ra1e6_comparison}    
\end{subfigure}    
    \caption{Rayleigh-Taylor results, IM = Implicit Midpoint, OL = One-Leg scheme.}
\label{fig:results_RayleighTaylor}
\end{figure}

{\begin{figure}[h]
\begin{subfigure}[b]{0.49\textwidth}    
  \centering
  \includegraphics[width=\textwidth]{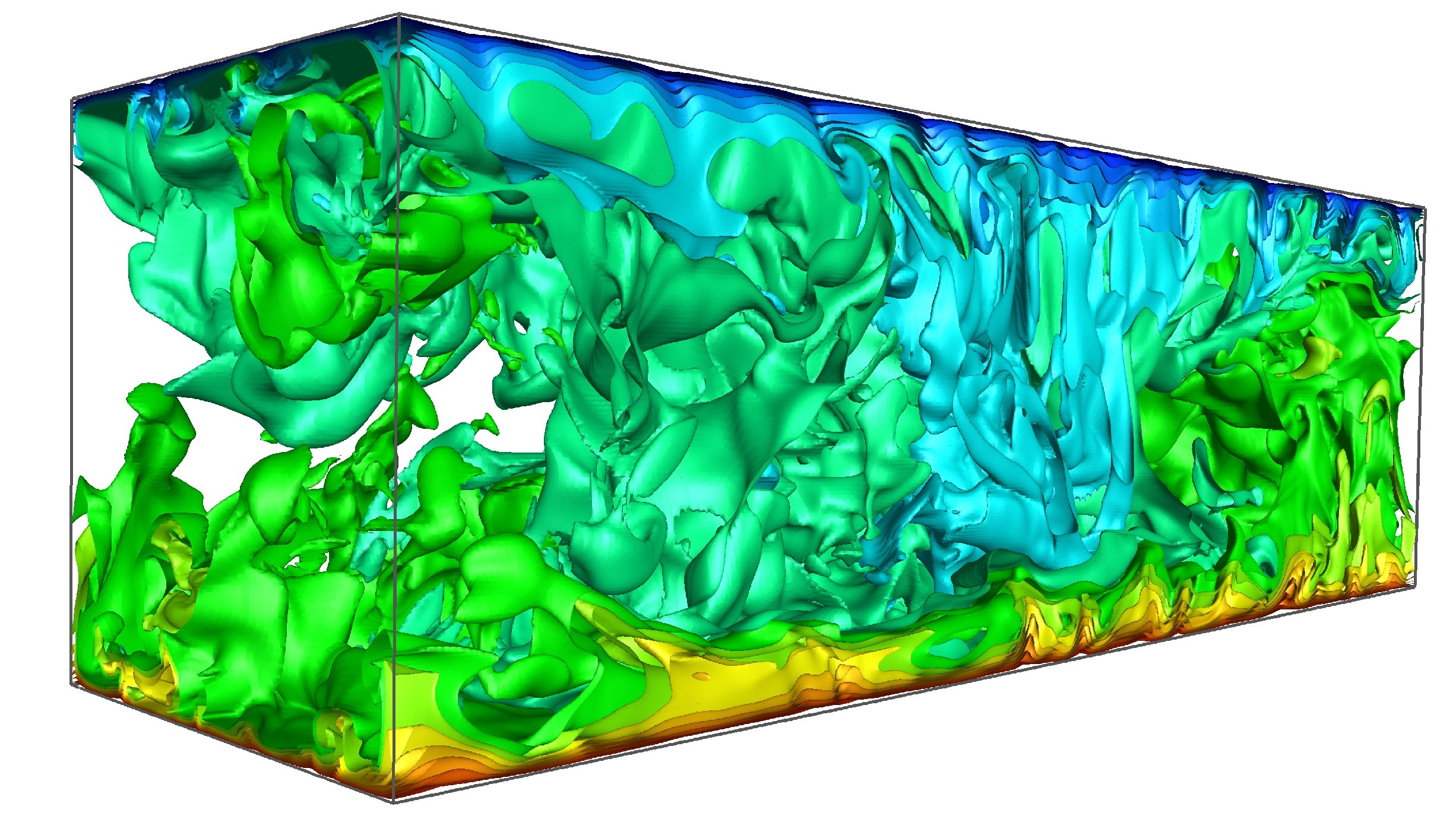}
  \caption{$\Ra=10^8$ and $\Ge=0$.}
  \label{fig:RBC_snap_Ra1e8_Ge0}
\end{subfigure}
\hfill
\begin{subfigure}[b]{0.49\textwidth}    
  \centering
  \includegraphics[width=\textwidth]{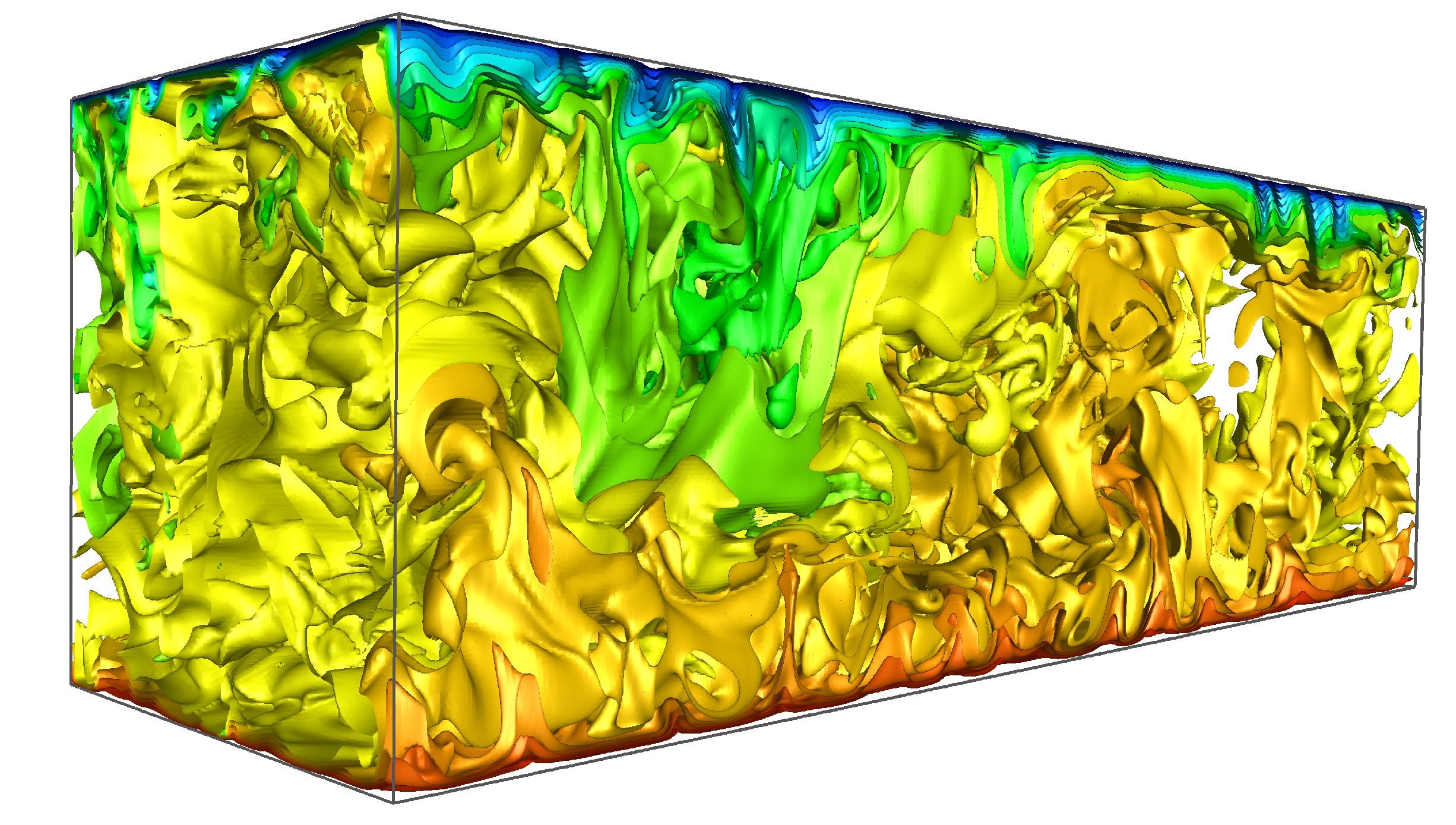}
  \caption{$\Ra=10^8$ and $\Ge=1$.}
  \label{fig:RBC_snap_Ra1e8_Ge1}
\end{subfigure}
\begin{subfigure}[b]{0.49\textwidth}    
  \centering
  \includegraphics[width=\textwidth]{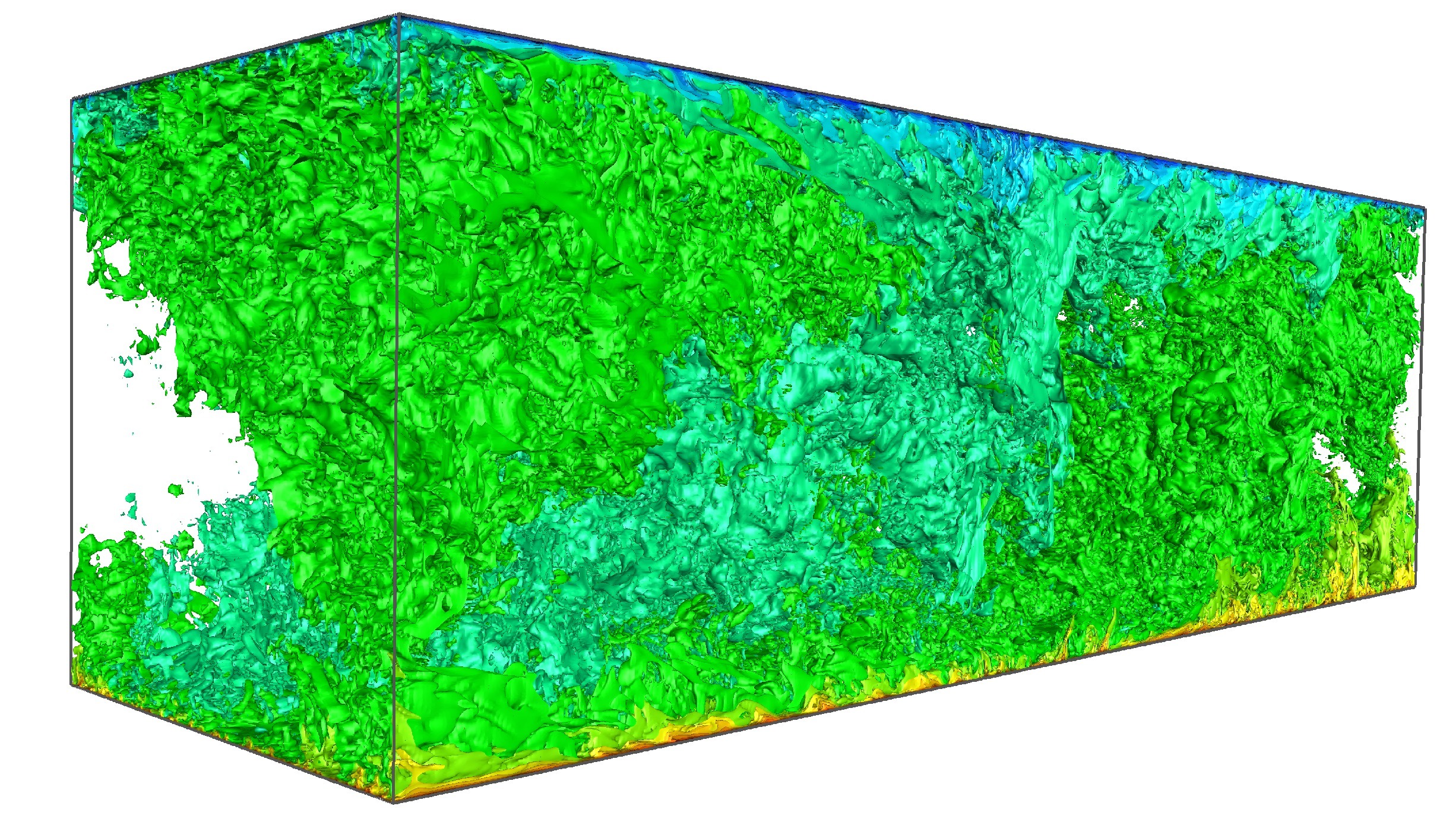}
  \caption{$\Ra=10^{10}$ and $\Ge=0$.}
  \label{fig:RBC_snap_Ra1e10_Ge0}
\end{subfigure}
\hfill
\begin{subfigure}[b]{0.49\textwidth}    
  \centering
  \includegraphics[width=\textwidth]{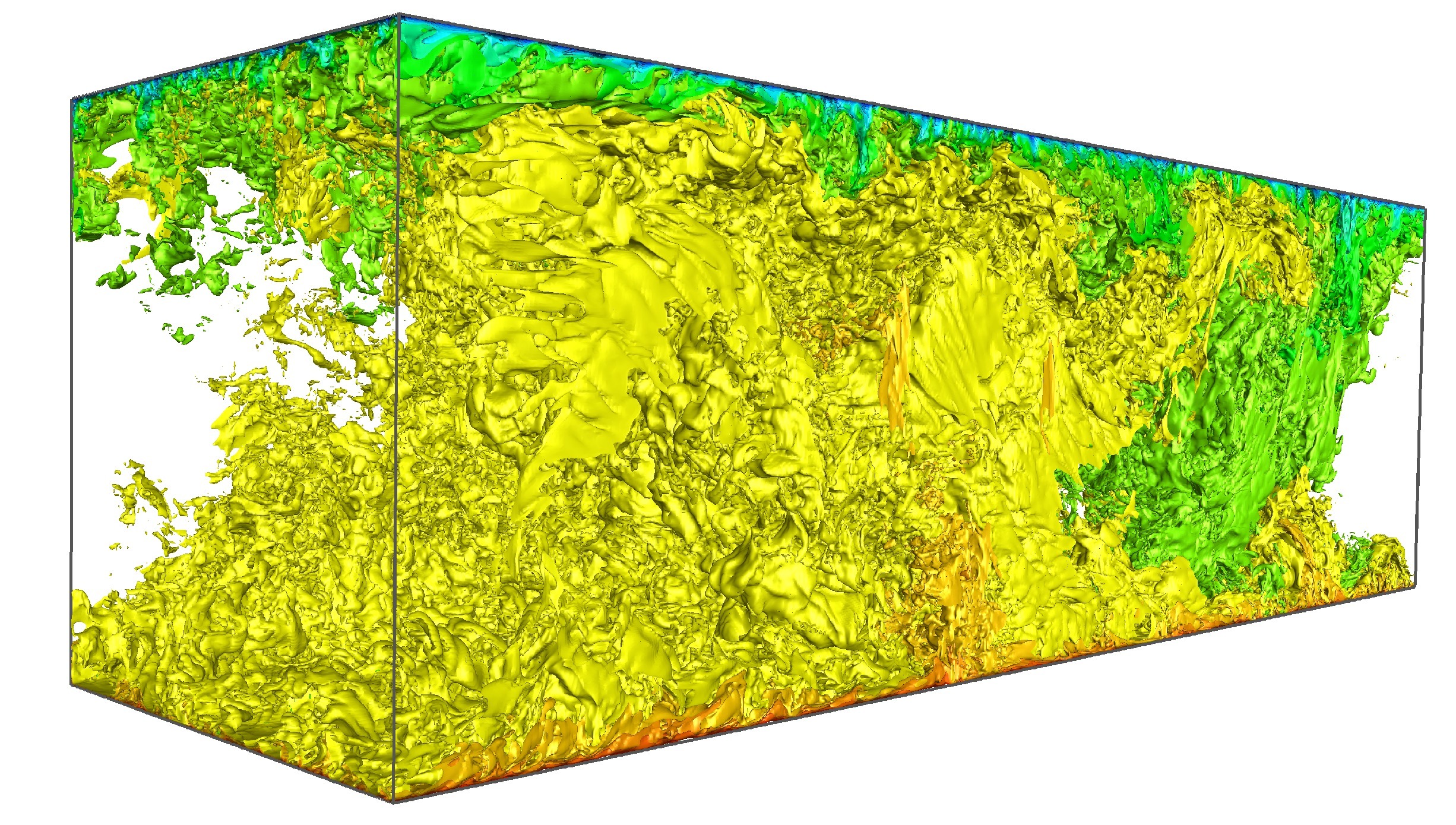}
  \caption{$\Ra=10^{10}$ and $\Ge=1$.}
  \label{fig:RBC_snap_Ra1e10_Ge1}
\end{subfigure}
\caption{Instantaneous temperature fields for 3D RBC at different Rayleigh and Gebhart numbers. For a visualization of the 3D time-dependent simulation results, we refer to the supplementary material.}
\label{fig:RBC_snaps}
\end{figure}

\section{Energy-conserving simulation of a turbulent flow}\label{sec:results_RBC3D}

As a final test-case, we consider the numerical simulation of an
air-filled ($\Pr=0.71$) Rayleigh--B\'{e}nard flow at two different
Rayleigh numbers, $\Ra=10^8$ and $10^{10}$. Direct numerical
simulations (DNS) were carried out and analyzed in previous
studies~\cite{DABTRI15-TOPO-RB,DABTRI19-3DTOPO-RB} without taking into
account the viscous dissipation effects ($\Ge=0$). Here, the results
are extended to $\Ge=0.1$ and $\Ge=1$ keeping the same domain size
($\pi \times 1 \times 1$) and mesh resolution ($400 \times 208 \times
208$ for $\Ra=10^8$, and $1024 \times 768 \times 768$ for
$\Ra=10^{10}$). Grids are constructed with a uniform grid spacing in
the periodic $x$-direction whereas wall-normal points ($y$ and $z$
directions) are distributed following a hyperbolic-tangent function as
follows (identical for the $z$-direction)
\begin{equation}
y_i = \frac{1}{2} \left( 1 + \frac{\tanh \left( \gamma_y ( 2 (i-1)/N_y - 1 ) \right) }{\tanh \gamma_y} \right) , \hspace{6mm} i=1,\dots,N_y+1,
\end{equation}
\noindent where $N_y$ and $\gamma_y$ are the number of control volumes
and the concentration factor in the $y$-direction, respectively. In
our case, $\gamma_y = \gamma_z = 1.4$ for $\Ra=10^8$ and $\gamma_y =
\gamma_z = 1.6$ for $\Ra=10^{10}$. For further details, the reader is
referred to our previous
works~\cite{DABTRI15-TOPO-RB,DABTRI19-3DTOPO-RB}.
% \bigskip

Instantaneous temperature fields corresponding to the statistically
steady state are displayed in Figure~\ref{fig:RBC_snaps}. As expected,
thermal dissipation effects at $\Ge=1$ lead to a significant increase
in the average cavity temperature which is clearly visible for both
Rayleigh numbers. As in 2D, the flow symmetry (in average sense)
with respect to the mid-height plane is lost for $\Ge > 0$ leading to
higher (lower) Nusselt number for the top (bottom) wall. Subsequently,
the top (bottom) thermal boundary layer becomes thinner (thicker)
with respect to the case at $\Ge=0$. This implies that mesh resolution
requirements in the near-wall region are also asymmetrical; however,
in this work, for the sake of simplicity, the grid spacing at the two walls is the same
regardless of the Gebhart number.

\begin{figure}[h]
  % \begin{subfigure}[b]{0.49\textwidth}
  %   \centering
  %   \includegraphics[width=\textwidth]
  %   {RBC1e8_budgets_Ge1_0_fine_with_dissipation.pdf}%RBC_energy_balances_Ra1e8_Ge1_fine
  %   \caption{Finest grid: $400 \times 208 \times 208 \approx 17.3\text{M}$.}
  %   \label{fig:RBC_enerbal_Ra1e8_Ge1_fine}
  % \end{subfigure}
  % \begin{subfigure}[b]{0.51\textwidth}
  %   \centering
  %   \includegraphics[width=\textwidth]{RBC1e8_budgets_Ge1_0_coarse_with_dissipation.pdf} %RBC_energy_balances_Ra1e8_Ge1_coarse
  %   \caption{Coarsest grid: $50 \times 26 \times 26 \approx 0.034\text{M}$.}
  %   \label{fig:RBC_enerbal_Ra1e8_Ge1_coarse}
  % \end{subfigure}
  \centering
  \includegraphics[width=\textwidth]{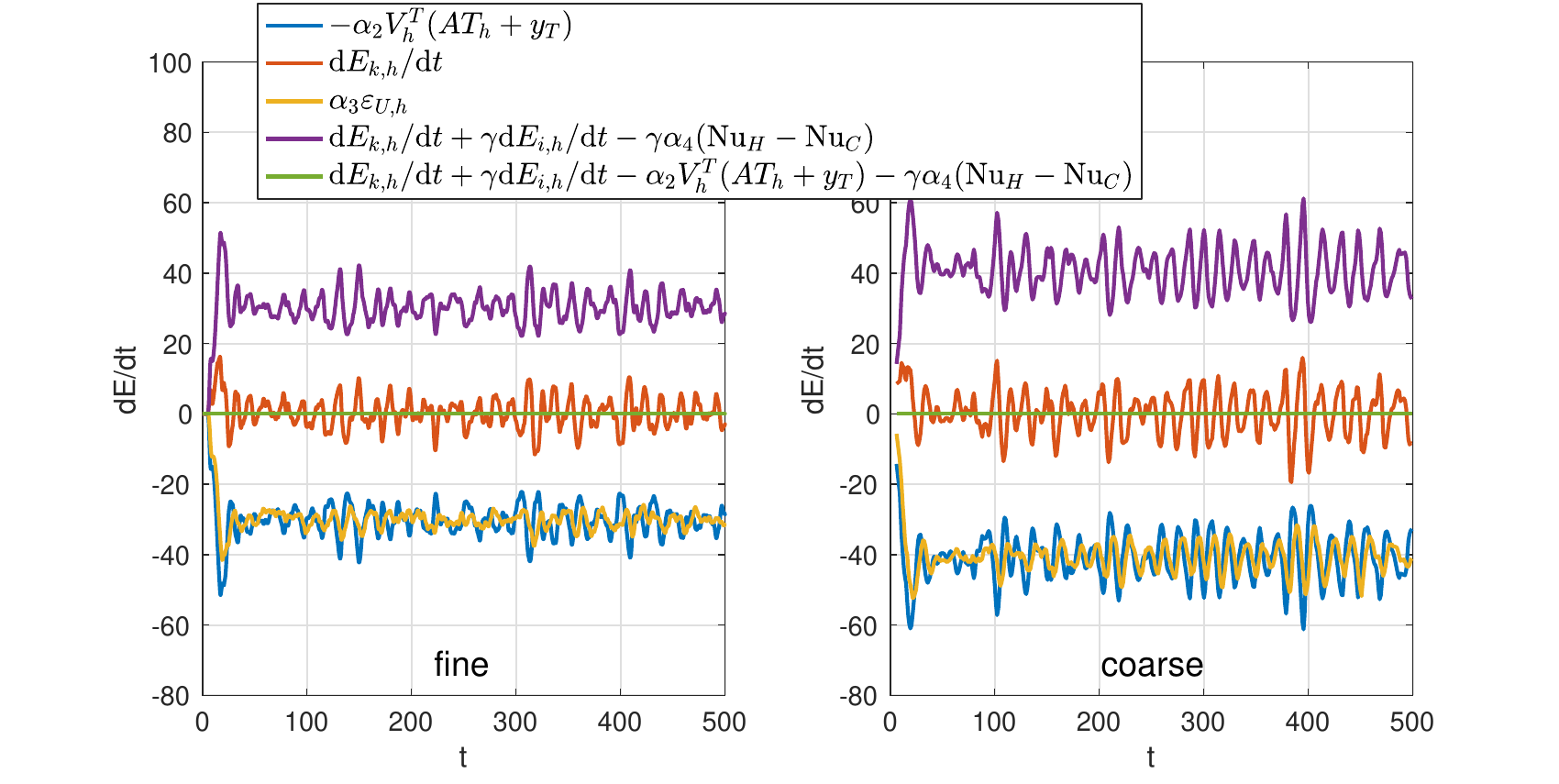}
\caption{Time-evolution of the most relevant energy contributions for $\Ge=1$; (left) Finest grid: $400 \times 208 \times 208 \approx 17.3\text{M}$; (right) Coarsest grid: $50 \times 26 \times 26 \approx 0.034\text{M}$.}
\label{fig:RBC_enerbal_Ra1e8_Ge1}
\end{figure}

All simulations have been carried out for $500$~time-units
starting from a zero velocity field and uniformly distributed random
temperatures between $T_C$ and $T_H$. As the fluid sets in motion, initially the
discrete kinetic energy of the system increases. Then, after a sufficiently long
period of time (around $50$~time-units) a statistically steady
state is reached. This is clearly observed in
Figure~\ref{fig:RBC_enerbal_Ra1e8_Ge1} where the time-evolution of
various rate-of-changes of energy are shown. Results correspond to
$\Ra=10^{8}$ and $\Ge=1$ using a very fine ($400 \times 208 \times 208
\approx 17.3\text{M}$) and a very coarse mesh. Similar results are
obtained for the other tested configurations. As expected, once a
statistically steady state is reached, the kinetic energy fluctuates
around its mean value and therefore its rate-of-change $\rd
E_{k,h}/\rd t$ (in red) fluctuates around zero. Only two terms contribute to
the global kinetic energy of the system (see
equation~\eqref{eqn:kineticenergy_semidiscrete}): the global viscous
dissipation, $\epsilon_{u,h}$ (in yellow), and the contribution of the buoyancy
forces given by $\alpha_{2} V_{h}^T (A T_{h}(t) + y_{T})$ (in blue). These two
contributions cancel each other on average when a statistically steady state is reached. The former is transferred into internal energy, $E_{i,h}$, whereas the latter can be
viewed as a transfer from potential to kinetic energy. In addition, the total energy of the system is exactly in balance with the buoyancy term and the heat conduction through the top and bottom boundaries (green line), as given by \eqref{eqn:totalenergy_balance_semidiscrete}, repeated here for convenience:
\begin{equation}
  \frac{\rd E_{k,h}}{\rd t} + \gamma \frac{\rd E_{i,h}}{\rd t} - \alpha_{2} V_{h}^T ( A T_{h} + y_{T} ) -  \gamma \alpha_{4} ( \Nu_H - \Nu_C ) = 0.
\end{equation}
This proofs that the viscous dissipation function has indeed been discretized correctly, since an imbalance between the viscous dissipation implied by the kinetic energy equation and the explicitly added viscous dissipation in the internal energy equation would otherwise show up. These energy balances are exactly satisfied for any grid, so even for very coarse grids (see Figure~\ref{fig:RBC_enerbal_Ra1e8_Ge1}, right). Notice again that it is important that the Nusselt numbers are evaluated consistently with the discretization of the diffusive terms in the internal energy equation, as explained in Section~\ref{sec:Nusselt_discretization}. 

%This is a highly relevant feature of the discretization approach presented in this work. Namely, energy transfers are exactly preserved at the discrete level without introducing any artificial source/sink of energy. 

% This is shown in Figure~\ref{fig:RBC_therbal} where (time-averaged) Nusselt numbers obtained for a wide range of meshes are displayed.

In addition to these instantaneous balances, we show in Figure~\ref{fig:RBC_therbal} that the time averages of the exact relations given in equations~\eqref{eqn:Nusselt_epsU}, \eqref{eqn:Nusselt_kinenergy} and~\eqref{eqn:Nusselt_thermal_diss} are preserved at the discrete level, similar to what was shown in steady-state in 2D (see Figure~\ref{fig:Ra_Nu_bifurcation_annotated}). 
However, here we display \textit{time-averaged} Nusselt numbers and consider a wide
range of meshes. The finest meshes correspond to the DNS
simulations shown in Figure~\ref{fig:RBC_snaps} whereas coarser and
coarser meshes have been generated by reducing the number of grid
points in each spatial direction by factors of approximately
$\sqrt{2}$. Hence, after six successive mesh coarsenings, the total
number of grid points is reduced by approximately $((\sqrt{2})^6)^3 =
2^9 = 512$. This under-resolution causes a pile-up of (kinetic) energy
close to the smallest resolved scales, that leads to higher values of
$\epsilon_U$ and, therefore, an increase of both $\Nu_H$ (see
equation~\eqref{eqn:Nusselt_kinenergy}) and $\Nu_C-\Nu_H$ (see
equation~\eqref{eqn:Nusselt_epsU}). 
Although the solution is surely less accurate at coarse grids, the fact that an energy balance is still satisfied, makes our approach an excellent starting point for developing or testing sub-grid scale models, as the additional dissipation that is introduced by the sub-grid scale model can be exactly quantified.
% , the solution errors can only be attributed to the (lack of) effect of the sub-grid scales. 
% Therefore, we consider that the discretization presented hereforms 
%
\begin{figure}[h]
  \begin{subfigure}[b]{0.49\textwidth}
    \centering
    \includegraphics[angle=0,width=\textwidth]{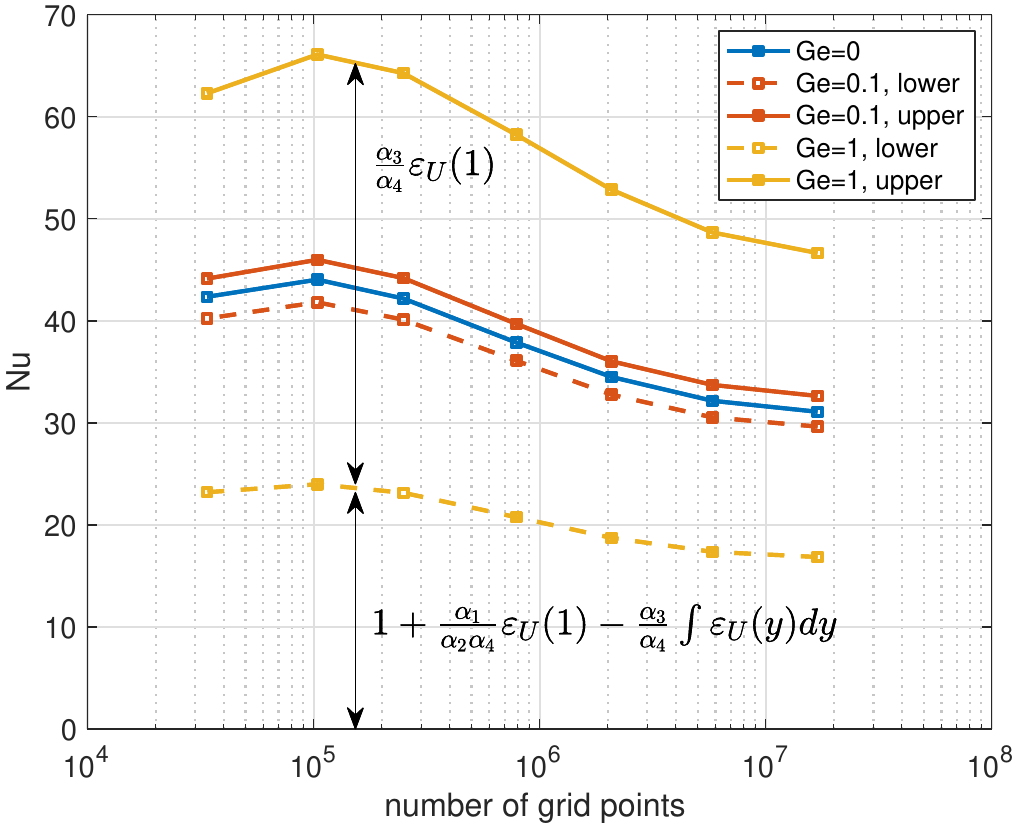} %thermal_budgets_Ra1e8.eps
    \caption{$\Ra=10^{8}$.}
    \label{fig:RBC_therbal_Ra1e8}
  \end{subfigure}
  \begin{subfigure}[b]{0.49\textwidth}
    \centering
    \includegraphics[angle=0,width=\textwidth]{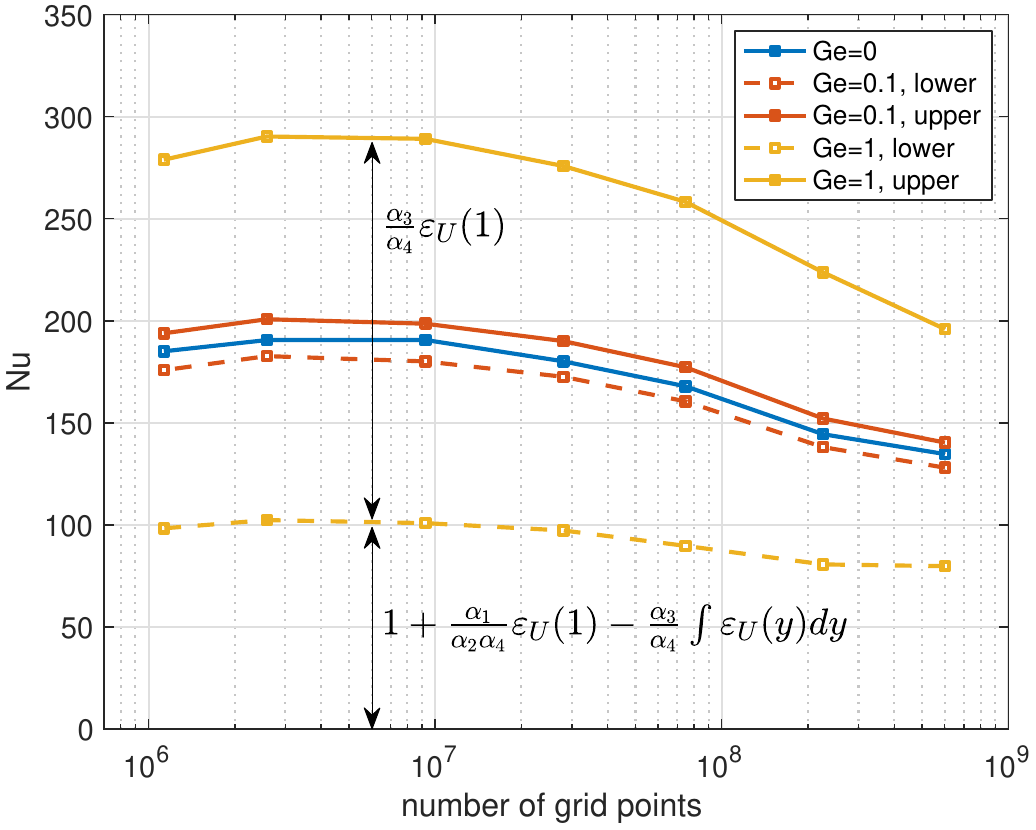}%thermal_budgets_Ra1e10.eps
    \caption{$\Ra=10^{10}$.}
    \label{fig:RBC_therbal_Ra1e10}
  \end{subfigure}
  \caption{Time-averaged Nusselt numbers at lower and upper plate for a set of meshes at $\Ge=0$, $\Ge=0.1$ and $\Ge=1$.}
  \label{fig:RBC_therbal}
\end{figure}

\section{Conclusions}
In this paper we have proposed a new energy-consistent discretization of the viscous dissipation function. The viscous dissipation function is an important quantity, for example in turbulent flow computations, where it is critical to assess the global energy balances, or in natural convection flows, where it leads to internal heating. This latter case has been the focus of this article, and we have shown that including the viscous dissipation function in the internal energy equation leads to a \textit{consistent total energy balance}: viscous dissipation acts as a sink in the kinetic energy equation and as a source in the internal energy equation, such that the sum of internal and kinetic energy only changes due to buoyancy and thermal diffusion.

Our key result is a new discretization of the \textit{local} viscous dissipation function that abides by the total energy balance. We have shown that it is determined by two choices, namely the discretization of the diffusive terms in the momentum equations and the expression for the local kinetic energy. The discretization of the diffusive terms is detailed for both general (non-constant viscosity) and simplified (constant viscosity) stress tensor expressions. The proposed expression for the local kinetic energy is such that a discrete local kinetic energy equation is satisfied, and leads to a quadratic, strictly dissipative form of the viscous dissipation function, also for general stress tensors. Near boundaries we have proposed corrections to the viscous dissipation function to keep the dissipative property.

The numerical experiments in 2D and 3D show that viscous dissipation does not affect the critical Rayleigh number at which instabilities form, but it does significantly impact the development of instabilities once they occur, leading to a significant difference between the Nusselt numbers on the cold and hot plates. Moreover, simulations of turbulent Rayleigh-B\'{e}nard convection have clearly shown that the proposed discretization is stable even for very coarse grids. Namely, the numerical discretization does not interfere with the energy balances and, therefore, we consider that the proposed method is an excellent starting point for testing sub-grid scale models.
 %solution errors can only be attributed to the lack of resolution. In this sense,

%In 3D, our results show a linear relation between the Nusselt number on the cold and hot plates, i.e.\ $\Nu_{C} = c \cdot \Nu_{H}$ with the constant of proportionality increasing with increasing Gebhart number. This relation could be useful in further extending the existing scaling theories for Rayleigh-Bénard convection.

The analysis in this paper has been performed for the classic finite-volume staggered grid method. Extensions to other discretization methods, such as finite differences or finite elements, are in principle possible provided that a discrete local kinetic energy balance mimicking the continuous balance can be identified. Another limitation of this work is the assumption of incompressible flow, which might seem restrictive given the fact that viscous dissipation typically becomes important for compressible flows. However, the idea of discretizing the viscous dissipation term in an energy-consistent manner is also applicable to compressible flows, see e.g.\ \cite{najm1998,nemati2016}, and we expect our work can therefore be extended in that direction.

As mentioned, an important avenue for future work lies in the assessment of subgrid-scale models for turbulent flows, including those driven by buoyancy. For example, in large-eddy simulation, the kinetic energy equation of the resolved scales and of the subgrid-scales features viscous dissipation terms, and the current work provides a basis for proper discrete representations of these terms.

\FloatBarrier

\section*{Data availability statement}
The incompressible Navier-Stokes code is available at \url{https://github.com/bsanderse/INS2D} (Matlab version). A Julia version is available from \url{https://github.com/agdestein/IncompressibleNavierStokes.jl}. The data generated in this work is available upon request.

\section*{CRediT}
\textbf{Benjamin Sanderse}: Conceptualization, Methodology, Writing - Original Draft, Software; \textbf{Xavier Trias}: Writing - Original Draft, Writing - Review \& Editing, Software

\section*{Acknowledgements}
This publication is part of the project "Discretize first, reduce next" (with project number VI.Vidi.193.105) of the research programme NWO Talent Programme Vidi which is (partly) financed by the Dutch Research Council (NWO). {F.X.T.} is supported by the \textit{Ministerio de Econom\'{i}a y Competitividad}, Spain, RETOtwin project (PDC2021-120970-I00). Turbulent Rayleigh-B\'{e}nard simulations were carried out on MareNostrum~4 supercomputer at BSC. The authors thankfully acknowledge these institutions.

\appendix

\section{Forms of the dissipation function}\label{sec:forms_dissipation_function}
In this appendix we expalin why the dissipation function changes depending on which form of the stress tensor is used. The stress tensor for an incompressible fluid with non-constant viscosity is given by
\begin{equation}
    \hat{\vt{\tau}} = \mu (\nabla \vt{u} + (\nabla \vt{u})^{T}).
\end{equation}
In the case of constant viscosity $\mu$, the divergence of the stress tensor can be simplified:
\begin{equation}
   \nabla \cdot \hat{\vt{\tau}} = \mu \nabla \cdot (\nabla \vt{u} + (\nabla \vt{u})^{T}) = \mu \nabla \cdot \nabla \vt{u} + \mu \nabla \cdot (\nabla \vt{u})^{T}= \mu \nabla^2 \vt{u} + \mu \nabla (\nabla \cdot \vt{u}) = \mu \nabla^2 \vt{u} =: \nabla \cdot \vt{\tau},
\end{equation}
where
\begin{equation}
    \vt{\tau} = \mu \nabla \vt{u} = \hat{\vt{\tau}} - \mu (\nabla \vt{u})^T.
\end{equation}
Note that $\vt{\tau}$ is not a proper stress tensor, since it is not symmetric. We stress that $\nabla \cdot \vt{\tau} = \nabla \cdot \hat{\vt{\tau}}$, even though $\vt{\tau} \neq \hat{\vt{\tau}}$.

In the kinetic energy equation the divergence of the stress tensor is multiplied by $\vt{u}$: $\vt{u} \cdot (\nabla \cdot \vt{\tau})$. Since $\nabla \cdot \vt{\tau} = \nabla \cdot \hat{\vt{\tau}}$, we also have 
\begin{equation}
    \vt{u} \cdot (\nabla \cdot \vt{\tau}) = \vt{u} \cdot (\nabla \cdot \hat{\vt{\tau}}).
\end{equation}
Expanding both the left-hand and right-hand side with a vector identity (note: also valid for non-symmetric $\vt{\tau}$) gives:
\begin{equation}
        \nabla \cdot (\vt{\tau} \cdot \vt{u}) - \vt{\tau} : \nabla \vt{u} =  \nabla \cdot (\hat{\vt{\tau}} \cdot \vt{u}) - \hat{\vt{\tau}} : \nabla \vt{u}. 
\end{equation}
or 
\begin{equation}\label{eqn:energy_diffusion}
        \nabla \cdot (\vt{\tau} \cdot \vt{u}) - \Phi =  \nabla \cdot (\hat{\vt{\tau}} \cdot \vt{u}) - \hat{\Phi}.
\end{equation}
The crucial point is that, even though equation \eqref{eqn:energy_diffusion} holds, the individual terms are not equal, i.e.\ $\Phi\neq \hat{\Phi}$ and $\nabla \cdot (\vt{\tau} \cdot \vt{u}) \neq \nabla \cdot (\hat{\vt{\tau}} \cdot \vt{u})$.

% Note that in terms of $\hat{\tau}$ and $\hat{\Phi}$ we can write the expression above as
% \begin{equation}
%     u_{i} \dd{\tau_{ij}}{x_j} = \dd{}{x_{j}} (\hat{\tau}_{ij} u_{i}) - \hat{\tau}_{ij} \dd{u_{i}}{x_{j}} =  \dd{}{x_{j}} (\hat{\tau}_{ij} u_{i}) - \hat{\Phi},
% \end{equation}
% (care should be taken because $\hat{\tau}$ is not symmetric) and this expression should equal $\dd{}{x_{j}} (\tau_{ij} u_{i})- \tau_{ij} \dd{u_{i}}{x_{j}}$.

 % We can investigate both forms by writing equation \eqref{eqn:tau_u} in Cartesian form as
This insight can be further clarified by evaluating these expressions in 2D Cartesian coordinates:
 \begin{align}    
    \nabla \cdot (\hat{\vt{\tau}} \cdot \vt{u})&= \mu \left[ \dd{}{x}\left(2u \dd{u}{x}\right) +  \dd{}{y} \left( u \left( \dd{u}{y} + \dd{v}{x} \right) \right) + \dd{}{x}\left(v \left( \dd{u}{y} + \dd{v}{x} \right)\right) +  \dd{}{y}\left(2 v\dd{v}{y}\right) \right], \\
     \hat{\Phi} &= \mu \left[ 2 \left( \dd{u}{x} \right)^2 + \left( \dd{u}{y} + \dd{v}{x} \right)^2 + 2 \left( \dd{v}{y} \right)^2 \right], \\
    \nabla \cdot (\vt{\tau} \cdot \vt{u})  &= \mu \left[ \dd{}{x}\left(u \dd{u}{x}\right) +  \dd{}{y} \left( u \dd{u}{y} \right) + \dd{}{x}\left(v \dd{v}{x} \right) + \dd{}{y}\left(v\dd{v}{y}\right) \right], \\
    \Phi &= \mu \left[ \left( \dd{u}{x} \right)^2 + \left( \dd{u}{y} \right)^2 + \left( \dd{v}{x} \right)^2 + \left( \dd{v}{y} \right)^2 \right].
\end{align}
Note that in a closed domain ($\vt{u}=0$ on all boundaries), we have the relation
\begin{equation}
    \int_{\Omega} \Phi \, \rd \Omega = \int_{\Omega} \hat{\Phi} \, \rd \Omega.
\end{equation}

{\section{Discrete dissipation operator from local kinetic energy equation}\label{sec:discrete_kinetic_energy}
% In order to derive an expression for $\Phi$, we need to consider the local kinetic energy equation.

\subsection{Momentum equations and choice of local kinetic energy}
The energy-conserving discretization presented in equation \eqref{eqn:mom_semidiscrete} can be written component-wise as:
% \begin{equation}\label{eqn:mom_discretization_u}
%     \frac{\rd u_{i+1/2,j}}{\rd t} + \text{conv}^{u}_{i+1/2,j} = -\frac{p_{i+1,j} - p_{i,j}}{\Delta x} + \mu \frac{u_{i+3/2,j} - 2 u_{i+1/2,j} + u_{i-1/2,j}}{\Delta x^2} + \mu \frac{u_{i+1/2,j+1} - 2 u_{i+1/2,j} + u_{i+1/2,j-1}}{\Delta y^2} 
% \end{equation}
\begin{align}\label{eqn:mom_discretization_u}
    \frac{\rd u_{i+1/2,j}}{\rd t} &= - \text{conv}^{u}_{i+1/2,j} -\frac{p_{i+1,j} - p_{i,j}}{\Delta x} + \alpha_{1} \text{diff}^{u}_{i+1/2,j}, \\
    \frac{\rd v_{i,j+1/2}}{\rd t} &= - \text{conv}^{v}_{i+1/2,j} -\frac{p_{i,j+1} - p_{i,j}}{\Delta y} + \alpha_{1} \text{diff}^{v}_{i,j+1/2} + \alpha_{2} \frac{1}{2}(T_{i,j}+T_{i,j+1}).
\end{align}
% \begin{equation}
%     \frac{\rd v_{i,j+1/2}}{\rd t} + \text{conv}^{v}_{i+1/2,j} = -\frac{p_{i,j+1} - p_{i,j}}{\Delta y} + \mu \frac{v_{i+1,j+1/2} - 2 v_{i,j+1/2} + v_{i-1,j+1/2}}{\Delta x^2} + \mu \frac{v_{i,j+3/2} - 2 v_{i,j+1/2} + v_{i,j-1/2}}{\Delta y^2} 
% \end{equation}
The convective terms are discretized starting from the divergence form, and due to discrete mass conservation they can be written in skew-symmetric form, which is energy-conserving. These terms are not the main focus of this work and we refer to \cite{verstappen2003,sanderse2013} for details.

The aim here is to find a local kinetic energy equation expression and the exact form of the dissipation terms. The local kinetic energy should be such that it results in the well-known global kinetic energy balance \cite{verstappen2003} upon integration over the entire domain. This  global kinetic energy equation is obtained by taking the inner product of all momentum equations with the full velocity vector $V_h$ (containing $u_{i+1/2,j}$ and $v_{i,j+1/2}$ at all locations). This resulting global kinetic energy definition $\frac{1}{2} V_{h}^{T} \Omega_{h} V_{h}$ still leaves room for the definition of the local kinetic energy.

Our proposal is to choose for the local kinetic energy the definition
\begin{equation}\label{eqn:local_KE_app}
    k_{i,j} := \frac{1}{4} u_{i+1/2,j}^2 + \frac{1}{4} u_{i-1/2,j}^2 + \frac{1}{4} v_{i,j+1/2}^2 + \frac{1}{4} v_{i,j-1/2}^2.
\end{equation}
Upon differentiating,
\begin{equation}\label{eqn:local_KE_ext}
    \frac{\rd k_{i,j}}{\rd t} = \frac{1}{2} u_{i+1/2,j} \frac{\rd u_{i+1/2,j} }{\rd t} + \frac{1}{2} u_{i-1/2,j} \frac{\rd u_{i-1/2,j} }{\rd t} + \frac{1}{2} v_{i,j+1/2} \frac{\rd v_{i+1/2,j} }{\rd t} + \frac{1}{2} v_{i,j-1/2} \frac{\rd v_{i,j-1/2} }{\rd t},
\end{equation}
and substituting the momentum equations, our proposed definition gives a local kinetic energy equation that is consistent with the continuous equations. The stencil of points required to evaluate \eqref{eqn:local_KE_ext} is shown in figure \ref{fig:energy_equation_stencil}.

\begingroup
\begin{figure}[hbtp]
\fontfamily{lmss} % Latin Modern Sans Serif
%\fontfamily{phv} % helvetica
\fontsize{11pt}{11pt}\selectfont
\centering 
\def\svgwidth{0.6 \textwidth}
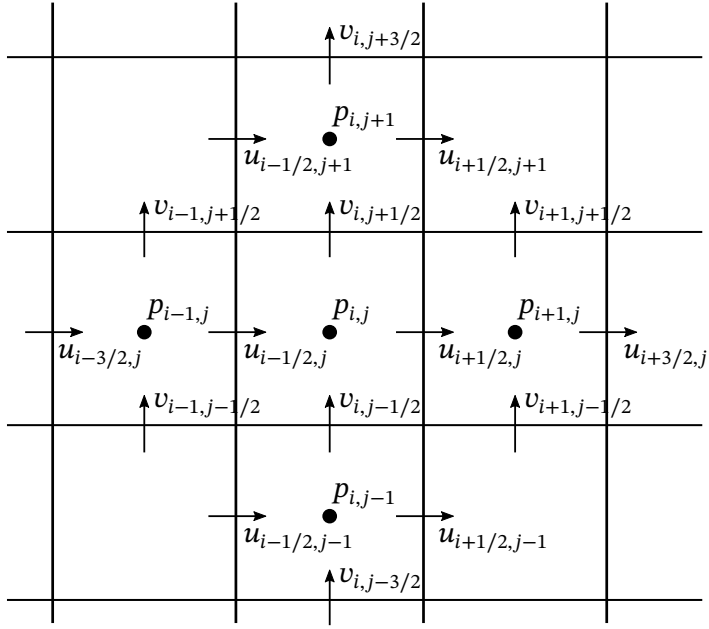 
\caption{Stencil of velocity and pressure points involved in the local kinetic energy equation.}
\label{fig:energy_equation_stencil}
\end{figure}
\endgroup

% \subsection{Pressure gradient}
The choice \eqref{eqn:local_KE_app} is inspired by the fact that it naturally allows a discrete equivalent of $\vt{u} \cdot \nabla p=\nabla \cdot (p \vt{u}) - p \nabla \cdot \vt{u}$:
% \subsection{Pressure gradient}
% We aim to find a discrete equivalent of $\vt{u} \cdot \nabla p = \nabla \cdot (p \vt{u}) - p \nabla \cdot \vt{u}$. Write:
\begin{multline}
    \frac{1}{2} u_{i+1/2,j} \frac{p_{i+1,j} - p_{i,j}}{\Delta x} + \frac{1}{2} u_{i-1/2,j} \frac{p_{i,j} - p_{i-1,j}}{\Delta x} + \frac{1}{2} v_{i,j+1/2} \frac{p_{i,j+1} - p_{i,j}}{\Delta y} + \frac{1}{2} v_{i,j-1/2} \frac{p_{i,j} - p_{i,j-1}}{\Delta y} = \\
        \frac{u_{i+1/2,j} \frac{1}{2} (p_{i+1,j} +p_{i,j})- \frac{1}{2} u_{i-1/2,j} (p_{i,j} + p_{i-1,j})}{\Delta x} + \frac{v_{i,j+1/2} \frac{1}{2} (p_{i,j+1} + p_{i,j}) - v_{i,j-1/2} \frac{1}{2} (p_{i,j} + p_{i,j-1})}{\Delta y}   \\
        - p_{i,j} \underbrace{\left(\frac{u_{i+1/2,j} -  u_{i-1/2,j}}{\Delta x} + \frac{v_{i,j+1/2} - v_{i,j-1/2}}{\Delta y} \right)}_{\text{div}( u)_{i,j}}.
    % = \frac{u_{i+1/2,j} + u_{i+3/2,j}}{2} p_{i+1,j} - \frac{u_{i-1/2,j} + u_{i+1/2,j}}{2} p_{i,j} \\ 
    % - \frac{1}{2} p_{i+1,j} (u_{i+3/2,j} - u_{i+1/2,j}) - \frac{1}{2} p_{i,j} (u_{i+1/2,j} - u_{i-1/2,j})
\end{multline}
Furthermore, choice \eqref{eqn:local_KE_app} for the local kinetic energy leads to a consistent quadratic dissipation form in the case of a general stress tensor, as will be shown in \ref{sec:KE_general_stress}.

\begin{comment}
\subsection{Convection}
\begin{equation}
C_{h}^{u} (V_{h},u_{h})_{i+1/2,j} :=  \bar{u}_{i+1,j} u_{i+1,j} - \bar{u}_{i,j} u_{i,j} + \bar{v}_{i+1/2,j+1/2} u_{i+1/2,j+1/2} -  \bar{v}_{i+1/2,j-1/2} u_{i+1/2,j-1/2}.
\end{equation}
\begin{multline}
\frac{1}{2} u_{i+1/2,j} \left[  \frac{1}{2} ( \bar{u}_{i+1/2,j} + \bar{u}_{i+3/2,j} ) -  \frac{1}{2} ( \bar{u}_{i-1/2,j} + \bar{u}_{i+1/2,j} ) +  \frac{1}{2} ( \bar{v}_{i,j+1/2} + \bar{v}_{i+1,j+1/2} ) -  \frac{1}{2} ( \bar{v}_{i,j-1/2} + \bar{v}_{i+1,j-1/2} ) \right] \\
+ \frac{1}{2} u_{i+3/2,j} \frac{1}{2} \left(\bar{u}_{i+1/2,j} + \bar{u}_{i+3/2,j} \right) - \frac{1}{2} u_{i-1/2,j} \frac{1}{2} \left(\bar{u}_{i-1/2,j} + \bar{u}_{i+1/2,j} \right) \\
 + \frac{1}{2} u_{i+1/2,j+1} \frac{1}{2} \left(\bar{v}_{i,j+1/2} + \bar{v}_{i+1,j+1/2} \right) - \frac{1}{2} u_{i+1/2,j-1} \frac{1}{2} \left(\bar{v}_{i,j-1/2} + \bar{v}_{i+1,j-1/2} \right),
\end{multline}

When adding $u_{i+1/2,j} \cdot (.)$ and $u_{i-1/2,j} \cdot (.)$ and $v_{i,j+1/2}$ etc., and using divergence-freeness, we get
\begin{equation}
   \frac{1}{4} u_{i+1/2,j} u_{i+3/2,j} \frac{1}{2} \left(\bar{u}_{i+1/2,j} + \bar{u}_{i+3/2,j} \right) - \frac{1}{4} u_{i-1/2,j} u_{i-3/2,j} \frac{1}{2} \left(\bar{u}_{i-3/2,j} + \bar{u}_{i-1/2,j} \right) + \ldots,
\end{equation}
which is the discrete approximation to $\dd{}{x} \cdot (\frac{1}{2} u^2 u)$.

\end{comment}

\subsection{Diffusion and dissipation}\label{sec:discrete_kinetic_energy_dissipation}
We continue to investigate the dissipation implied by the diffusive term in the momentum equation \eqref{eqn:mom_discretization_u} and the kinetic energy choice \eqref{eqn:local_KE_app}. 
Restricting ourselves momentarily to the term $\frac{\partial^2 u}{\partial x^2}$, we are looking for a discrete equivalent of the relation
\begin{equation}\label{eqn:dissipation_1D}
     u \frac{\partial^2 u}{\partial x^2} =  - \left( \frac{\partial u}{\partial x}\right)^2 + \frac{\partial}{\partial x} \left( u \frac{\partial u}{\partial x}\right).
\end{equation}
  % in the dissipation caused by such a term in the kinetic and internal energy equations,
% Consider the second-order diffusion stencil
% \begin{equation}
%     \frac{\rd^2 u}{\rd x^2} \approx \frac{u_{i+3/2} - 2 u_{i+1/2} + u_{i-1/2}}{\Delta x^2} = \frac{1}{\Delta x} \left( \frac{u_{i+3/2} - u_{i+1/2}}{\Delta x} - \frac{u_{i+1/2} - u_{i-1/2}}{\Delta x}  \right)
% \end{equation}
This is given by $u_{i+1/2,j} \cdot \text{diff}^{u}_{i+1/2,j}$:
\begin{multline}\label{eqn:discrete_local_dissipation}
    \frac{u_{i+1/2,j}}{\Delta x} \left( \frac{u_{i+3/2,j} - u_{i+1/2,j}}{\Delta x} - \frac{u_{i+1/2,j} - u_{i-1/2,j}}{\Delta x}  \right) 
    =   - \frac{1}{2} \left( \frac{u_{i+3/2,j} - u_{i+1/2,j}}{\Delta x} \right)^2 - \frac{1}{2} \left( \frac{u_{i+1/2,j} - u_{i-1/2,j}}{\Delta x} \right)^2\\
    + \frac{1}{\Delta x} \left( \frac{1}{2}(u_{i+3/2,j} + u_{i+1/2,j})\frac{u_{i+3/2,j} - u_{i+1/2,j}}{\Delta x} - \frac{1}{2}(u_{i+1/2,j} + u_{i-1/2,j})\frac{u_{i+1/2,j} - u_{i-1/2,j}}{\Delta x} \right) .
\end{multline}
% Upon summation over a periodic domain, the first term on the rhs cancels (telescoping property), and we can write
% \begin{equation}
%     u_{h}^T D_{h} u_{h} = -\sum_{i} \left( \frac{u_{i+1/2} - u_{i-1/2}}{\Delta x} \right)^2,
% \end{equation}
% and note that the factor $\frac{1}{2}$ disappears.
Equation \eqref{eqn:discrete_local_dissipation} is important because the discrete local dissipation expression is explicitly needed in the internal energy equation.

The analysis for the term $\frac{\partial^2 u}{\partial y^2}$ is completely analogous, and hence we can define the following discrete function that describes the dissipation implied by the discretized diffusion term of the momentum equation for $u_{i+1/2,j}$:
\begin{multline}
    \Phi^{u}_{i+1/2,j} = - \frac{1}{2} \left( \frac{u_{i+3/2,j} - u_{i+1/2,j}}{\Delta x} \right)^2 -   \frac{1}{2} \left( \frac{u_{i+1/2,j} - u_{i-1/2,j}}{\Delta x} \right)^2  \\ -\frac{1}{2}  \left( \frac{u_{i+1/2,j+1} - u_{i+1/2,j}}{\Delta y} \right)^2 - \frac{1}{2} \left( \frac{u_{i+1/2,j} - u_{i+1/2,j-1}}{\Delta y} \right)^2.
\end{multline}
Similarly, the dissipation implied by the discretized diffusion term of the momentum equation for $v_{i,j+1/2}$ is:
\begin{multline}
    \Phi^{v}_{i,j+1/2} = - \frac{1}{2}  \left( \frac{v_{i+1,j+1/2} - v_{i,j+1/2}}{\Delta x} \right)^2  - \frac{1}{2} \left( \frac{v_{i+1,j-1/2} - v_{i,j-1/2}}{\Delta x} \right)^2  \\ - \frac{1}{2} \left( \frac{v_{i,j+3/2} - v_{i,j+1/2}}{\Delta y} \right)^2 - \frac{1}{2}  \left( \frac{v_{i,j+1/2} - v_{i,j-1/2}}{\Delta y} \right)^2.
\end{multline}
The entire dissipation term appearing in the kinetic energy equation for $\frac{\rd k_{i,j}}{\rd t}$ is then
\begin{equation}\label{eqn:local_dissipation_function_appendix}
    \boxed{\Phi_{i,j} = \frac{1}{2} \Phi^{u}_{i+1/2,j} + \frac{1}{2} \Phi^{u}_{i-1/2,j} + \frac{1}{2} \Phi^{v}_{i,j+1/2} + \frac{1}{2} \Phi^{v}_{i,j-1/2}.}
\end{equation}

\subsection{Boundary conditions}\label{sec:dissipation_BC}
The analysis in the previous section ignored the effect of boundary conditions. Upon integrating \eqref{eqn:dissipation_1D} over the domain, we get
\begin{equation}
    \int u \frac{\partial^2 u}{\partial x^2} \rd x = - \int \left( \frac{\partial u}{\partial x}\right)^2 \rd x  + \underbrace{\left[ u \frac{\partial u}{\partial x} \right]}_{\text{boundary term}},
\end{equation}
and the boundary term vanishes in case of homogeneous Dirichlet, homogeneous Neumann or periodic conditions. The discrete version should mimic this behavior. 

\begingroup
\begin{figure}[hbtp]
\fontfamily{lmss} % Latin Modern Sans Serif
%\fontfamily{phv} % helvetica
\fontsize{11pt}{11pt}\selectfont
\centering 
\def\svgwidth{ \textwidth}
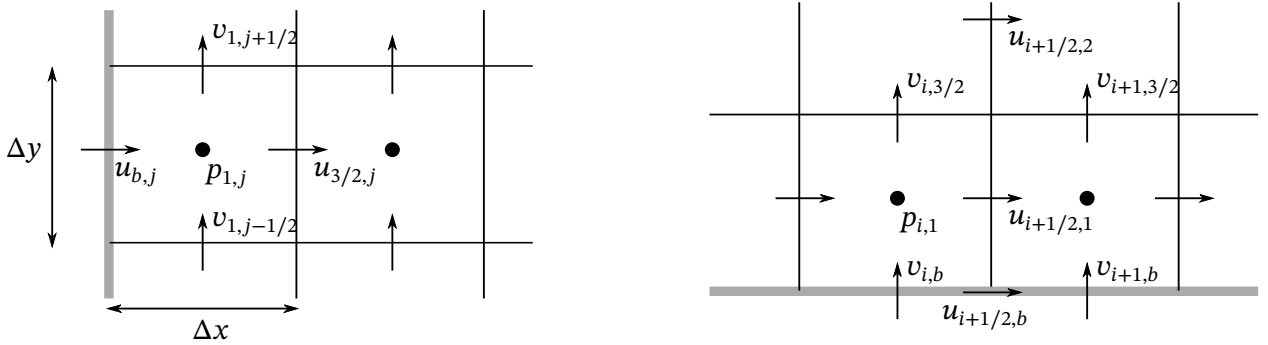 
\caption{Staggered grid near vertical (left) and horizontal (right) boundary.}
\label{fig:staggered_boundary}
\end{figure}
\endgroup

Consider the case where the solution on the boundary is given by $u_{1/2,j} = u_{b,j}$ (figure \ref{fig:staggered_boundary}, left). Then the first unknown for which the momentum equation \eqref{eqn:mom_discretization_u} is solved is $u_{3/2,j}$, and equation \eqref{eqn:discrete_local_dissipation} becomes
\begin{multline}\label{eqn:discrete_local_dissipation_BC}
    \frac{u_{3/2,j}}{\Delta x} \left( \frac{u_{5/2,j} - u_{3/2,j}}{\Delta x} - \frac{u_{3/2,j} - u_{b,j}}{\Delta x}  \right)   = 
    \frac{1}{\Delta x} \left( \frac{1}{2}(u_{5/2,j} + u_{3/2,j})\frac{u_{5/2,j} - u_{3/2,j}}{\Delta x} - \frac{1}{2}(u_{3/2,j} + u_{b,j})\frac{u_{3/2,j} - u_{b,j}}{\Delta x} \right) \\
    - \frac{1}{2} \left( \frac{u_{5/2,j} - u_{3/2,j}}{\Delta x} \right)^2 - \frac{1}{2} \left( \frac{u_{3/2,j} - u_{b,j}}{\Delta x} \right)^2.
\end{multline}
In case where $u_{b,j}=0$, we want the boundary term to vanish, like the term $u \frac{\partial u}{\partial x}$ in the continuous case. However, when setting $u_{b,j}=0$, the term that corresponds to $u \frac{\partial u}{\partial x}$ reads:
\begin{equation}
    -\frac{1}{2}(u_{3/2,j} + u_{b,j})\frac{u_{3/2,j} - u_{b,j}}{\Delta x} = -\frac{1}{2} \frac{u_{3/2,j}^2}{\Delta x},
\end{equation}
and the discrete boundary contribution does \textit{not} vanish for $u_{b,j}=0$. 
% This is an issue, which we resolve by realizing that $-\frac{1}{2} \frac{u_{3/2}^2}{\Delta x}$ is always negative and can be absorbed into the definition of $\Phi$ near the boundary. In other words, we define
% \begin{equation}
%     \tilde{\Phi}_{1,j} = \Phi_{1,j} - \frac{1}{2} \frac{u_{3/2,j}^2}{\Delta x}.
% \end{equation}
% With this definition we get $V_{h}^T D_{h} V_{h} = \sum_{i,j} \tilde{\Phi}_{i,j} \Omega_{i,j}$.
% Another option is to write
This issue is caused by the fact that the finite volumes do not cover the entire domain, because there is no momentum equation to be solved for $u_{b,j}$ (as it is given by the boundary data). We resolve this issue by splitting instead as
\begin{equation}
    - \frac{u_{3/2,j}}{\Delta x} \left(\frac{u_{3/2,j} - u_{b,j}}{\Delta x} \right) = - \underbrace{\frac{u_{b,j}}{\Delta x} \left(\frac{u_{3/2,j} - u_{b,j}}{\Delta x} \right)}_{\textrm{boundary contribution}} - \underbrace{\frac{u_{3/2,j} - u_{b,j}}{\Delta x} \left( \frac{u_{3/2,j} - u_{b,j}}{\Delta x} \right)}_{\textrm{dissipation contribution}},
\end{equation}
so that the contribution to the dissipation function is 
\begin{equation}\label{eqn:Phi_BC}
    - \left(\frac{u_{3/2,j} - u_b,j}{\Delta x}\right)^2,
\end{equation}
instead of $ - \frac{1}{2} \left(\frac{u_{3/2,j} - u_{b,j}}{\Delta x}\right)^2$.
% which is a more general formulation in case $u_b \neq 0$.

For the discretization of $\frac{\partial^2 u}{\partial y^2}$, we have a different situation, because the solution points are not aligned with the boundary. The first unknown is $u_{i+1/2,1}$, which is situated at a distance $\frac{1}{2} \Delta y$ above the lower boundary. In this case we can write
% In case the solution point does \textit{not} coincide with the boundary, as is for example the case when approximating $\frac{\rd^2 u}{\rd y^2}$ at the first point $j=1$ by $(u_{i+1/2,j+1} - 2 u_{i+1/2,j} + u_{i+1/2,j-1})/\Delta y^2 = (u_{i+1/2,2} - 3 u_{i+1/2,1} + 2 u_{b})/\Delta y^2$, this issue does \textit{not} occur, and no correction to $\Phi$ is needed. This is seen from rewriting 
% \begin{equation}
%     \begin{split}
\begin{multline}
        \frac{u_{i+1/2,j}}{\Delta y}\left(\frac{u_{i+1/2,j+1} - u_{i+1/2,j}}{\Delta y} - \frac{u_{i+1/2,j} - u_{i+1/2,j-1}}{\Delta y} \right) \overset{j=1}{=} \frac{u_{i+1/2,1}}{\Delta y} \left(\frac{u_{i+1/2,2} - u_{i+1/2,1}}{\Delta y} - \frac{u_{i+1/2,1} - u_{i+1/2,b}}{\frac{1}{2} \Delta y}\right) \\
        = \frac{1}{\Delta y}\left( \frac{1}{2}(u_{i+1/2,2} + u_{i+1/2,1}) \frac{u_{i+1/2,2} - u_{i+1/2,1}}{\Delta y}  - u_{i+1/2,b}\frac{u_{i+1/2,2} - u_{i+1/2,1}}{\frac{1}{2}\Delta y} \right) \\
        -\frac{1}{2}\left(\frac{u_{i+1/2,2} - u_{i+1/2,1}}{\Delta y}\right)^2  - \frac{1}{2} \left(\frac{u_{i+1/2,1} - u_{i+1/2,b}}{\frac{1}{2} \Delta y}\right)^2,
\end{multline}
and we have a correct discrete equivalent of the continuous expression, and no correction to $\Phi$ is needed.

The analysis for the $v$-component follows in a similar fashion. A correction is needed in the expression for $\Phi$ associated to $\frac{\partial^2 v}{\partial x^2}$, but not for $\frac{\partial^2 v}{\partial y^2}$.

\subsection{Extension to non-constant viscosity: general stress tensor}\label{sec:KE_general_stress}
For the case of non-constant $\mu$, the discretization of the diffusion terms in the momentum equation changes to
\begin{equation}\label{eqn:diff_u_nonconstant_mu}
\begin{split}
 \text{diff}^{u}_{i+1/2,j} =&    \frac{1}{\Delta x} \left[  \left(2\mu_{i+1,j} \frac{u_{i+3/2,j} - u_{i+1/2,j}}{\Delta x}\right) - \left(2\mu_{i,j} \frac{u_{i+1/2,j} - u_{i-1/2,j}}{\Delta x}\right) \right] + \\ 
    &\frac{1}{\Delta y} \left[  \left(\mu_{i+1/2,j+1/2} \frac{u_{i+1/2,j+1} - u_{i+1/2,j}}{\Delta y}\right) - \left(\mu_{i+1/2,j-1/2} \frac{u_{i+1/2,j} - u_{i+1/2,j-1}}{\Delta y}\right) \right] +\\
    & \frac{1}{\Delta y} \left[ \left(\mu_{i+1/2,j+1/2} \frac{v_{i+1,j+1/2} - v_{i,j+1/2}}{\Delta x}\right) - \left(\mu_{i+1/2,j-1/2} \frac{v_{i+1,j-1/2} - v_{i,j-1/2}}{\Delta x}\right) \right]. 
\end{split}
\end{equation}
Importantly, we first show that this form reduces to the expression in equation \eqref{eqn:mom_discretization_u} for constant $\mu$. In the continuous equations, this happens because 
\begin{equation}
    \dd{}{x} \left(2 \dd{u}{x} \right) + \dd{}{y} \left(\dd{u}{y} + \dd{v}{x}\right) = \dd{}{x} \left(\dd{u}{x} \right) + \dd{}{y} \left(\dd{u}{y}\right) + \dd{}{x}\left(\dd{u}{x} + \dd{v}{y}\right) = \dd{^2u}{x^2} + \dd{^2 u}{y^2}.
\end{equation}
The derivation hinges on the divergence-freeness of $\vt{u}$ and interchanging of differentiation in $x-$ and $y$-directions. Discretely, the same derivation holds, which can be shown by rewriting as follows:
\begin{equation}
\begin{split}
 % \text{diff}_{u,i+1/2,j} =
 &    \frac{1}{\Delta x} \left[  \left(\frac{u_{i+3/2,j} - u_{i+1/2,j}}{\Delta x}\right) - \left(\frac{u_{i+1/2,j} - u_{i-1/2,j}}{\Delta x}\right) \right] +
    \frac{1}{\Delta y} \left[  \left( \frac{u_{i+1/2,j+1} - u_{i+1/2,j}}{\Delta y}\right) - \left(\frac{u_{i+1/2,j} - u_{i+1/2,j-1}}{\Delta y}\right) \right] +\\
  & \frac{1}{\Delta x} \left[  \left(\frac{u_{i+3/2,j} - u_{i+1/2,j}}{\Delta x}\right) - \left(\frac{u_{i+1/2,j} - u_{i-1/2,j}}{\Delta x}\right) + \left( \frac{v_{i+1,j+1/2} - v_{i+1,j-1/2}}{\Delta y}\right) - \left(\frac{v_{i,j+1/2} - v_{i,j-1/2}}{\Delta y}\right) \right], 
\end{split}
\end{equation}
and the second line evaluates to zero, as it contains the difference of the divergence associated to volumes $(i+1,j)$ and $(i,j)$.

We continue to derive the dissipation function. As explain in Remark \ref{sec:remark_dissipation_form} and in \ref{sec:forms_dissipation_function}, the dissipation function changes when the generic stress tensor for non-constant $\mu$ is considered. Multiplying \eqref{eqn:diff_u_nonconstant_mu} with $u_{i+1/2,j}$ and rewriting leads to
\begin{multline}
    \hat{\Phi}^{u}_{i+1/2,j} = u_{i+1/2,j} \cdot \text{diff}^{u}_{i+1/2,j} = \\
    \frac{1}{\Delta x} \left( \mu_{i+1,j} (u_{i+3/2,j} + u_{i+1/2,j}) \frac{u_{i+3/2,j} - u_{i+1/2,j}}{\Delta x} - \mu_{i,j} (u_{i+1/2,j} + u_{i-1/2,j}) \frac{u_{i+1/2,j} - u_{i-1/2,j}}{\Delta x} \right) \\
    - \mu_{i+1,j} \left( \frac{u_{i+3/2,j} - u_{i+1/2,j}}{\Delta x} \right)^2 - \mu_{i,j} \left( \frac{u_{i+1/2,j} - u_{i-1/2,j}}{\Delta x} \right)^2 +\\
    \frac{1}{\Delta y} \Bigg( \mu_{i+1/2,j+1/2} \frac{u_{i+1/2,j+1} + u_{i+1/2,j}}{2} \left[\frac{u_{i+1/2,j+1} - u_{i+1/2,j}}{\Delta y} + \frac{v_{i+1,j+1/2} - v_{i,j+1/2}}{\Delta x} \right] \\
    - \mu_{i+1/2,j-1/2} \frac{u_{i+1/2,j} + u_{i+1/2,j-1}}{2} \left[ \frac{u_{i+1/2,j} - u_{i+1/2,j-1}}{\Delta y} + \frac{v_{i+1,j-1/2} - v_{i,j-1/2}}{\Delta x} \right] \Bigg) \\
    - \frac{\mu_{i+1/2,j+1/2}}{2} \left( \frac{u_{i+1/2,j+1} - u_{i+1/2,j}}{\Delta y} \right) \left( \frac{u_{i+1/2,j+1} - u_{i+1/2,j}}{\Delta y} + \frac{v_{i+1,j+1/2} - v_{i,j+1/2}}{\Delta x} \right) \\
    - \frac{\mu_{i+1/2,j-1/2}}{2} \left( \frac{u_{i+1/2,j} - u_{i+1/2,j-1}}{\Delta y} \right) \left( \frac{u_{i+1/2,j} - u_{i+1/2,j-1}}{\Delta y} + \frac{v_{i+1,j-1/2} - v_{i,j-1/2}}{\Delta x} \right).
\end{multline}
The last two terms are not in quadratic form yet. The quadratic form results upon considering the full kinetic energy expression \eqref{eqn:local_KE_app}, i.e.\ adding $\hat{\Phi}^{u}_{i-1/2,j}= u_{i-1/2,j} \cdot \text{diff}^{u}_{i-1/2,j}$, $\hat{\Phi}^{v}_{i,j+1/2} = v_{i,j+1/2} \cdot \text{diff}^{v}_{i,j+1/2}$ and $\hat{\Phi}^{v}_{i,j-1/2} = v_{i,j-1/2} \cdot \text{diff}^{v}_{i,j-1/2}$. The full dissipation function then reads
\begin{multline}\label{eqn:Phi_hat}
    \hat{\Phi}_{i,j} = \frac{1}{2}\hat{\Phi}^{u}_{i+1/2,j} + \frac{1}{2} \hat{\Phi}^{u}_{i-1/2,j} + \frac{1}{2} \hat{\Phi}^{v}_{i,j+1/2} +\frac{1}{2} \hat{\Phi}^{v}_{i,j-1/2} = \\ 
    -\frac{\mu_{i+1,j}}{2} \left( \frac{u_{i+3/2,j} - u_{i+1/2,j}}{\Delta x} \right)^2 - \mu_{i,j} \left( \frac{u_{i+1/2,j} - u_{i-1/2,j}}{\Delta x} \right)^2 - \frac{\mu_{i-1,j}}{2} \left( \frac{u_{i-1/2,j} - u_{i-3/2,j}}{\Delta x} \right)^2 \\
    - \frac{\mu_{i+1/2,j+1/2}}{4} \left( \frac{u_{i+1/2,j+1} - u_{i+1/2,j}}{\Delta y} + \frac{v_{i+1,j+1/2} - v_{i,j+1/2}}{\Delta x} \right)^2
    - \frac{\mu_{i+1/2,j-1/2}}{4} \left( \frac{u_{i+1/2,j} - u_{i+1/2,j-1}}{\Delta y} + \frac{v_{i+1,j-1/2} - v_{i,j-1/2}}{\Delta x} \right)^2 \\
     - \frac{\mu_{i-1/2,j+1/2}}{4} \left( \frac{u_{i-1/2,j+1} - u_{i-1/2,j}}{\Delta y} + \frac{v_{i,j+1/2} - v_{i-1,j+1/2}}{\Delta x} \right)^2
    - \frac{\mu_{i-1/2,j-1/2}}{4} \left( \frac{u_{i-1/2,j} - u_{i-1/2,j-1}}{\Delta y} + \frac{v_{i,j-1/2} - v_{i-1,j-1/2}}{\Delta x} \right)^2 \\
     -\frac{\mu_{i,j+1}}{2} \left( \frac{v_{i,j+3/2} - v_{i,j+1/2}}{\Delta y} \right)^2 - \mu_{i,j} \left( \frac{v_{i,j+1/2} - v_{i,j-1/2}}{\Delta y} \right)^2 - \frac{\mu_{i,j-1}}{2} \left( \frac{v_{i,j-1/2} - v_{i,j-3/2}}{\Delta y} \right)^2.
\end{multline}

\bibliographystyle{elsarticle-num}
\bibliography{library_zotero.bib,mybiblio.bib}

\end{document}